\documentclass[structabstract]{aa}  
\usepackage{graphicx}
\usepackage{txfonts}

\usepackage{natbib}
\bibpunct{(}{)}{;}{a}{}{,} 

\begin{document}

%

\newcommand{\eeee}{El.}
\newcommand{\ione}{{\,\sc i}}
\newcommand{\itwo}{{\,\sc ii}}
\newcommand{\loggfvald}{$\log gf$ (VALD)}
\newcommand{\loggfcorr}{$\log gf$ (Corr.)}

\newcommand{\vmicro}{$\xi_t$}
\newcommand{\meh}{[M/H]}
\newcommand{\zet}{$Z$}
\newcommand{\vsini}{$v \sin i$}
\newcommand{\hbeta}{H$_\beta$}

\newcommand{\sep}{$\Delta \nu$}
\newcommand{\teff}{$T_{\rm eff}$}
\newcommand{\teffsun}{$T_{\rm eff; \odot}$}

\newcommand{\loggf}{$\log gf$}
\newcommand{\logg}{$\log g$}
\newcommand{\logghr}{$\log g_\nu$}
\newcommand{\loggnu}{$\log g_{\rm HR}$}
\newcommand{\loggstr}{$\log g_{m_1}$}

\newcommand{\feh}{[Fe/H]}
\newcommand{\kms}{km\,s$^{-1}$}
\newcommand{\mhz}{$\mu$Hz}
\newcommand{\iet}{{\sc i}}
\newcommand{\ito}{{\sc ii}}
\newcommand{\itre}{{\sc iii}}
\newcommand{\ifire}{{\sc iv}}

\newcommand{\lumn}{$L/{\rm L}_\odot$}
\newcommand{\mass}{$M/{\rm M}_\odot$}
\newcommand{\radius}{$R/{\rm R}_\odot$}

\newcommand{\shotgun}{{\sc shotgun}}

\newcommand{\atlasni}{ATLAS9}
\newcommand{\marcs}{MARCS}
\newcommand{\vwa}{VWA}
\newcommand{\eg}{e.g.}
\newcommand{\ie}{i.e.}
\newcommand{\cf}{cf.}
\newcommand{\rms}{RMS}

\newcommand{\templogg}{TEMPLOGG}
\newcommand{\synth}{SYNTH}
\newcommand{\simbad}{SIMBAD}
\newcommand{\vald}{VALD}
\newcommand{\hipp}{\emph {Hipparcos}}
\newcommand{\str}{Str\"omgren}
\newcommand{\corot}{{\em CoRoT}}
\newcommand{\pavo}{PAVO}
\newcommand{\chara}{CHARA}
\newcommand{\narval}{NARVAL}
\newcommand{\elodie}{ELODIE}
\newcommand{\harps}{HARPS}
\newcommand{\amber}{AMBER}
\newcommand{\vlti}{VLTI}

%

   \title{Accurate fundamental parameters of \corot\ asteroseismic targets}
\subtitle{The solar-like stars HD~49933, HD~175726, HD~181420 and HD~181906}

   \author{H. Bruntt
          \inst{1}
          }
   \institute{{Observatoire de Paris, LESIA, 5 place Jules Janssen, 92195 Meudon Cedex, France}
              \email{bruntt@phys.au.dk}
              \and
              {Sydney Institute for Astronomy, School of Physics, University of Sydney, 
               NSW 2006, Australia}
          }
   \date{Received February XXXX, 2009; accepted July YYYY, 2009}

 
  \abstract
   {The \corot\ satellite has provided 
high-quality light curves of several solar-like stars.
Analysis of the light curves provides oscillation frequencies 
that make it possible to probe the interior of the stars. However,
additional constraints on the fundamental parameters of the stars
are important for the theoretical modelling to be successful.
   }
   {We will estimate the fundamental parameters (mass, radius and luminosity)
of the first four solar-like targets to be observed in the asteroseismic field.
In addition, we will determine their effective temperature, 
metallicity and detailed abundance pattern.
   }
   {To constrain the stellar mass, radius and age we use
the \shotgun\ software which compares the location of the stars in 
the Hertzsprung-Russell diagram with theoretical evolution models. 
This method takes into account the uncertainties of the observed
parameters including the large separation determined
from the solar-like oscillations.
We determine the effective temperatures and abundance patterns 
in the stars from the analysis of high-resolution spectra. 
   }
   {We have determined the mass, radius and luminosity 
of the four \corot\ targets to within 
$5$--$10$\%, $2$--$4$\% and $5$--$13$\%, respectively. 
The quality of the stellar spectra determines how well
we can constrain the effective temperature. 
For the two best spectra we get $1$-$\sigma$ uncertainties below 60\,K and for
the other two $100$--$150$\,K.   
The uncertainty on the surface gravity is less than $0.08$\,dex for
three stars while for HD~181906 it is $0.15$\,dex.
The reason for the larger uncertainty is that the spectrum has two components
with a luminosity ratio of $L_{\rm p}/L_{\rm s} = 0.50\pm0.15$.
While \hipp\ astrometric data strongly suggest it is a binary star we
find evidence that the fainter star may be a background star,
since it is less luminous but hotter.
}
   {}




   \maketitle
%


\section{Introduction}

One of the ultimate science goals of the asteroseismic investigation
of the \corot\ mission is to compare the observed oscillation frequencies
with theoretical pulsation models. This will in principle allow us to
probe the inside of stars and in particular determine how well we think
we understand the physics of stars. We can examine how well the models
describe the observations and hopefully be able to improve some of the
approximations that are necessary when computing theoretical models.

To be able to confront the theoretical models with the observations
we need accurate and reliable estimates of the fundamental parameters of the stars,
\ie\ mass, radius and luminosity.
In this present work we will discuss how these parameters can be
estimated by comparison with theoretical evolution models.
In addition, we determine the atmospheric parameters from
photometric indices and detailed spectroscopic
analysis of absorption lines to get the chemical composition.
In particular, we will give results
for four solar-like stars observed in the asteroseismic field 
of \corot: HD~49933, HD~175726, HD~181420 and HD~181906.

%
%
\begin{table}
\caption{Properties of the observed spectra. 
The first column gives the abbreviation used for each spectrum.
Also listed is the resolution of the spectrograph, 
the wavelength range used in the abundance analysis,
the S/N in the continuum for the original resolution (column 5)
and S/N$_{42}$ for a resolution of $42\,000$. 
The last column contains the measured \vsini.
\label{tab:spec}}
%
%
\centering          
\setlength{\tabcolsep}{3pt}
\begin{tabular}{r@{}l l|rccrc}


\hline\hline       
               &         & Spectro- &                             &  &     &            & \vsini \\
HD&        /Sp.  & graph    &  \multicolumn{1}{c}{$R$} & Range [\AA] & S/N & S/N$_{42}$ & [\kms]  \\ 
\hline 

 49933&/H-A  & \harps   & 115\,000 & 4050--6850 & 400 & 740  &  10  \\
 $ $  &/H-C  & \harps   & 115\,000 & 4050--6850 & 430 & 790  & $-$  \\ \hline
175726&/N    & \narval  &  65\,000 & 4150--8500 & 900 &1110  &  12  \\ 
 $ $  &/E    & \elodie  &  42\,000 & 4050--6800 & 130 & 130  & $-$  \\ \hline 
181420&/F    & FEROS    &  48\,000 & 3800--8850 & 120 & 120  &  18  \\ 
 $ $  &/E    & \elodie  &  42\,000 & 4300--6800 & 120 & 120  & $-$  \\ \hline
181906&/E    & \elodie  &  42\,000 & 4450--6750 & 110 & 110  &  10  \\ 

\hline                    
\hline                  
\end{tabular}
\end{table}



\section{Observations}

For the target HD~49933 we used spectra from a 10-day
spectroscopic campaign with the \harps\ spectrograph \citep{mosser05}. 
From their bisector analysis it was found
that stellar activity was affecting the cores of the spectral
lines and this varied from night to night.
We stacked $49$ spectra from the quiet period 
(labeled ``H-C'' in the following) and 
collected 2001 January 19--20 from UT-22h to UT-06h. 
We stacked $35$ from a more active period (labeled ``H-A'')
collected the following night from UT-20h to UT-05h. 
The H-A and H-C spectra have signal-to-noise ratios (S/N) 
in the continuum of just over 400.


Spectra of HD~175726, HD~181420 and HD~181906 were obtained 
from the GAUDI database \citep{solano05}.
These spectra were collected as preparation for the \corot\ mission 
with the \elodie\ spectrograph 
mounted on the 1.93-m telescope at the Haute-Provence observatory.
The typical S/N in these spectra is 120.
Additional spectra of HD~175726 were obtained with the \narval\
spectrograph on the 2m~Bernard Lyot Telescope at the Pic~du~Midi
observatory as part of the \corot\ follow-up program.
We stacked 13 spectra to get a S/N of 900.
A spectrum of HD~181420 was also obtained with the FEROS spectrograph at ESO.

\begin{figure}
   \centering
\hskip -0.4cm
   \includegraphics[angle=90,width=9.4cm]{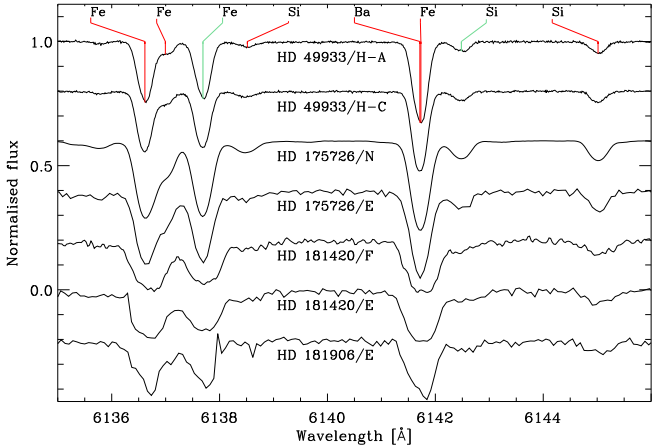}
   \caption{A section of the observed spectra for the four stars (\cf\ Table~\ref{tab:spec}).
The absorption lines due to Fe, Si and Ba are marked.
              \label{fig:spec}}
    \end{figure}

A small section of the observed spectra is shown in Fig.~\ref{fig:spec}
and the general properties of the spectra are listed in Table~\ref{tab:spec}.
The S/N was measured in the continuum as 
the average of several regions in the range $5\,500$--$6\,500$\,\AA.
We calculated S/N for the original resolution 
and when degraded to $R=42\,000$ (S/N$_{42}$ in Table~\ref{tab:spec})
with two pixels per resolution element.

We determined the projected rotation velocity (\vsini\,) 
by fitting synthetic spectra to several isolated
lines while assuming a macroturbulence of 2\,\kms\ for all stars.
The uncertainty on \vsini\ is 5\% except for HD~181906 which 
has two components in the spectrum, so the uncertainty 
is larger, $v \sin i=10\pm1$\,\kms. 
\cite{nordstrom04} determined a much higher 
value of $v \sin i = 16$\,\kms, 
which we also get if we assume it is a single star.
We shall discuss the possible binary nature of HD~181906 in Sect.~\ref{sec:bin}.
For the other three stars our \vsini\ values agree 
with \cite{nordstrom04} within the uncertainty,
although we note that their values are all exactly 1\,\kms\ higher.
 

%
%
%
%
%

\section{Fundamental parameters of stars}

The fundamental parameters of a star are its 
mass, $M$, radius, $R$, and luminosity, $L$.
A direct measurement of mass can be done for the components
in detached eclipsing binary systems with an accuracy 
of about 1--2 per cent \citep{andersen91}
or for single stars in the rare case of a microlensing event \citep{alcock01}.
The radius of single stars can be measured to about 1 per cent 
with interferometric methods or to about the same precision 
for eclipsing binary stars. The luminosity can be
estimated from the $V$ magnitude using the parallax and 
a bolometric correction, which is determined from model atmospheres.


For the relatively faint \corot\ asteroseismic targets
we cannot apply these direct methods to get the mass and radius,
so we must rely on indirect methods.
An advantage for the solar-like \corot\ targets is that
their density can be constrained from the large separation (\sep)
which is measured to 2 per cent or better \citep{michel08}
from the oscillation frequencies.

More detailed investigations involve comparison of 
the individual observed frequencies with theoretical models.
The aim of this work is to estimate the parameters that are 
used by the theoreticians, 
namely $L$, effective temperature, \teff, surface gravity, \logg,
and the chemical composition.
In this Section we shall discuss how the fundamental
parameters of stars can be determined.


%
 %
 %

    \begin{figure}
    \centering
    \includegraphics[width=8.0cm]{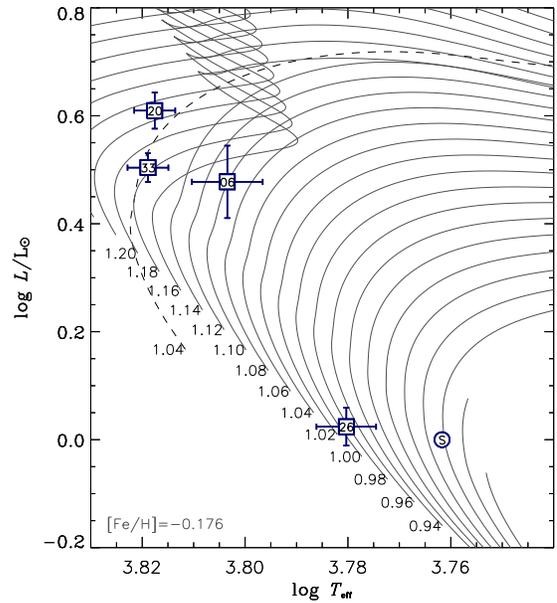}
    \caption{Hertzsprung-Russell diagram showing the position of the four \corot\ stars.
Full lines are evolution tracks from ASTEC for $[{\rm Fe/H}] = -0.176$ 
and the dashed line is for ${\rm [Fe/H]}=-0.480$.
The uncertainties and location of each target is marked by 
a box and identified by the last two digits in the HD number.
The mass is indicated for each track and the location of the Sun is marked with ``S''.
               \label{fig:hr}}
     \end{figure}
 %
 %

\subsection{Surface gravity from $M$ and $R$}
\label{sec:mass}

The position of the four \corot\ targets are shown 
in the Hertzsprung-Russell (H-R) diagram in Fig.~\ref{fig:hr}, 
where ASTEC evolution tracks from \cite{jcd08} are overlaid. 
The position of the Sun is shown for reference.
The mass and radius of a single field star can be 
estimated by comparing its properties to 
such theoretical evolution tracks. 

We adopted the so-called \shotgun\ method
which is described in detail in \cite{stello09}.
In brief, we construct 200 values for the four parameters (\teff, \lumn, \feh, \sep)
with a mean value equal to our estimated values (see below) and a 
Gaussian random scatter equal to the uncertainty.
For each of the 200 values we locate the closest matching 
grid point among the evolution tracks and the method saves 
the mass, radius and age. We finally calculate the mean value 
and the \rms\ value as our estimate of the $1\,\sigma$ uncertainty. 
The method uses the large separation to constrain the 
fundamental properties of the targets, and we assume that it
scales with the Sun as 
$\Delta \nu / \Delta \nu_\odot = \sqrt{\rho / \rho_\odot}$,
following the arguments by \cite{kjeldsen95}.
The inclusion of the large separation constrains 
the possible radii of the models as was discussed in detail by \cite{stello09}.

Compared to the description of \shotgun\ in \cite{stello09} 
we have increased the number of random seed values from 
100 to 200 and we have used both BASTI isochrones \citep{basti04} and 
ASTEC evolution tracks \cite{jcd08}. 
We find that the results for the four targets agree within 
the $1$-$\sigma$ uncertainty for both sets of models.
In the following we quote the results using the ASTEC evolution 
tracks, for which we have a denser model grid.

The four input parameters to \shotgun\ are estimated from:
\begin{itemize}

\item \teff: The $V-K$ index (calibration from \citealt{masana06}).

\item \lumn: $V$ from SIMBAD, \hipp\ parallax \citep{leeuwen07}, 
and bolometric correction \citep{bessell98}.

\item \feh: Initial value from the \str\ $m_1$ index and the 
final value from the abundance analysis.

\item \sep: Results from \corot\ except for HD~175726. 

\end{itemize}

For the \sep\ values we used the most resent results from \corot.
\cite{app08} analysed a 60-day light curve from \corot\ using 
different techniques and found $\Delta \nu = 85.9\pm0.15$\,\mhz.
\cite{michel08} presented preliminary results of the analysis of 
HD~181420 and HD~181906 and found $\Delta \nu = 77\pm2$ and $88\pm2$\,\mhz, respectively.
Finally, for HD~175726 we used 
$\Delta \nu = 97.2\pm0.5$\,\mhz\ from \cite{mosser09}.
This indicates a quite evolved star which is
not in agreement with the location of the star in the H-R diagram.
The detailed analysis of the oscillation frequencies show 
that the large separation is modulated with 
frequency \citep{mosser09},
so one should be cautious when using the scaling relation for \sep.
We therefore decided not to use the \sep\ value as input to \shotgun\
for HD~175726.

In Table~\ref{tab:fund1} we list the fundamental parameters from \shotgun.
The $1$-$\sigma$ uncertainty on the mass
is $5$--$10$\%, radius $2$--$4$\% and luminosity to $5$--$13$\%. 
The surface gravity is computed as $\log g = \log (M/{\rm M}_{\odot}) - 2 \log (R / {\rm R}_{\odot}) + \log g_\odot$,
and the results are listed in Table~\ref{tab:comp}.
The formal $1$-$\sigma$ uncertainty on \logg\ is 0.06 dex or 
slightly less for the four targets.
We note that our estimates of the uncertainties 
do not include the effects of changing the input physics of the models,
\eg\ mixing length, the equation of state or element diffusion.
However, the more detailed investigation of \cite{stello09} shows that
at least the radius is not significantly affected by changing these parameters.

%
%
%
\begin{table}
\caption{Fundamental parameters found from photometric indices.
\label{tab:phot2}}

\centering          
\setlength{\tabcolsep}{4pt}
\begin{tabular}{r|r@{}r|ccc}
\hline\hline       
                        & \multicolumn{2}{c|}{$V-K$} & \multicolumn{3}{|c}{\str\ indices} \\ 
\multicolumn{1}{c|}{HD} & \teff                      &    & \teff & \logg & \feh \\
\hline 



 49933 &  $6590\pm$&$ 60$ & $6630\pm90$ & $4.30\pm0.15$ & $-0.48\pm0.12$ \\
175726 &  $6030\pm$&$ 80$ & $5980\pm90$ & $4.68\pm0.15$ & $-0.21\pm0.10$ \\
181420 &  $6570\pm$&$ 60$ & $6660\pm90$ & $4.16\pm0.15$ & $-0.07\pm0.11$ \\
181906 &  $6360\pm$&$100$ & $6430\pm90$ & $4.31\pm0.15$ & $-0.20\pm0.11$ \\


\hline                    
 \hline                  
\end{tabular}
\end{table}
%

\subsection{The effective temperature}
\label{sec:teff}

The effective temperature of a star, \teff, is defined 
from the total flux per unit area from a black body:
$\sigma\,T_{\rm eff}^4 = F_{\rm tot} =\int_0^\infty F_\nu \, d\nu =  L / 4 \pi R^2$.
To make a direct determination of \teff\ one must measure
$F_{\rm tot}$ from the angular diameter (from interferometry)
and the flux integrated over all wavelengths (from spectrophotometry).
This has been done recently for a few nearby 
solar-like asteroseismic targets 
(see \citealt{north07,north09}) 
and yield \teff\ to about 50\,K. 
Interferometric measurements of the relatively faint \corot\ targets ($V\simeq5$--$8$)
may be possible in the near future with the recently installed
\pavo\ instrument at the \chara\ array \citep{ireland08} or \amber\ at \vlti\ \citep{amber07}.



When such measurements are not available, 
\teff\ can be estimated by indirect methods, \eg\ 
using the Balmer lines, 
calibration of line depth ratios,
photometric colour indices and detailed abundance analyses (see Sect.~\ref{sec:vwa}).
We did not use the Balmer lines due to the difficulty of normalising these
lines that cover several spectral orders. 
The calibration of line-depth rations of late F--K type stars 
was done by \cite{kovtyukh03}, but the valid range is 4000--6150\,K, 
and could thus only be used for the coolest star in our sample, 
HD~175726 (see \citealt{kovtyukh04}).
Several calibrations of photometric indices exist in the literature, \eg\ for
F- and G-type main sequence stars the $V-K$ index \citep{masana06} and
\str\ indices \citep{ramirez05} have recently been calibrated.
The $V-K$ index has the advantage that it is less sensitive 
to interstellar reddening and in the following we have adopted 
the \teff\ obtained from this index as our initial value 
for \shotgun\ and the abundance analyses.

%
\begin{table}
\caption{Fundamental parameters of the \corot\ targets. 
The age, mass and radius was found using the \shotgun\ method.
The luminosity is determined from $V$, the bolometric correction and the parallax.
\label{tab:fund1}}
\centering          
\setlength{\tabcolsep}{4pt}
\begin{tabular}{r| r@{}l cc l@{}l} 
\hline\hline       
\multicolumn{1}{c|}{HD} & 
             \multicolumn{2}{c}{Age [Gyr]} &$M/{\rm M}_\odot$& $R/{\rm R}_\odot$ &
                                                        \multicolumn{2}{c}{$L/{\rm L}_\odot$} \\ \hline 
 49933                 & $4.4$  &$\pm1.0$  & $1.079\pm0.073$ &  $1.385\pm0.031$  & $3.47 $&$\pm0.18 $ \\ 
175726                 & $4.8$  &$\pm3.5$  & $0.993\pm0.060$ &  $1.014\pm0.035$  & $1.210$&$\pm0.064$ \\ 
181420                 & $2.7$  &$\pm0.4$  & $1.311\pm0.063$ &  $1.595\pm0.032$  & $4.28 $&$\pm0.28 $ \\ 
181906                 & $4.2$  &$\pm1.6$  & $1.144\pm0.119$ &  $1.392\pm0.054$  & $3.29 $&$\pm0.43 $ \\ 
\hline 


\end{tabular}
\end{table}

We used the \str\ indexes from \cite{hauck98} 
as input to the \templogg\ \citep{rogers95, kupka01}
software\footnote{On line version: {\em http://www.univie.ac.at/asap/templogg/main.php}} 
to investigate if any of the stars have significant interstellar reddening. 
This requires the \hbeta\ index, which has not been measured for HD~175726.
For HD~49933, HD~181420 and HD~181906 we find slightly 
{\em negative} values for $E(b-y)$,
although they are consistent with zero within the uncertainties.
The distance to the four \corot\ stars range from 26--68\,parsec 
and in the following we assume that the targets 
have interstellar reddening close to zero, \ie\ $E(b-y)=0.000\pm0.005$. 
This corresponds to an uncertainty on \teff\ of $50$\,K
which we include in our estimates of \teff\ from the photometric indices.


In Table~\ref{tab:phot2} we give the results using the photometric indices.
We used 2MASS $V-K$ from \cite{cutri03} 
to get \teff\ using the calibration of \cite{masana06}.
We also list \teff\ using the \str\ calibration of \cite{ramirez05}.
There is a very good agreement on the \teff\ determined from the photometric indices.
The \teff\ determined from line depth ratios for HD~175726 is 
$6036\pm15$\,K (\citealt{kovtyukh04}; internal error is quoted) 
and this is fully consistent with the photometric calibrations.
In Table~\ref{tab:phot2} we also list \logg\ from \templogg\ 
and \feh\ using the \str\ calibration from \cite{martell02} (see below).

%
%

\subsection{Metallicity}
\label{sec:feh}

The metallicity of F- and G-type stars 
has been calibrated through the \str\ $m_1$ index,
which measures the strength of metal lines in a narrow band around 400--420\,nm.
{The calibration has recently been revised by \cite{martell02} and 
the 1-$\sigma$ scatter of their calibration is $0.09$\,dex. To this we add
the internal scatter assuming that each photometric index ($b-y$, $c_1$, $m_1$)
has a 1-$\sigma$ uncertainty of $0.005$.
The results are given in the last column in Table~\ref{tab:phot2}.

A more detailed view of the composition of stars is found from 
abundance analysis of individual spectral absorption lines formed in
the photosphere of the star. 
It is important to realise that we can only assume 
that this represents the global chemical composition of the star.
Diffusion processes and element levitation are important in some stars
and will change the radial distribution of elements. 
When modelling stars the chemical composition is often 
parametrised in terms of the mass fractions of hydrogen ($X$), 
helium ($Y$) and the rest is ``heavy elements'' ($Z$).
From spectroscopy of solar-type stars we cannot constrain the helium content.
Carbon, nitrogen and oxygen each give a significant contribution to $Z$.
Therefore, it is important to obtain spectra that cover the
near infrared where some carbon and oxygen lines are found 
(\ie\ the spectra we have from FEROS and \narval).

The results of abundance analyses rely on the adopted model atmospheres
that depend on \teff, \logg, \feh, and microturbulence (\vmicro).
A detailed analysis requires a good spectrum, say ${\rm S/N} > 150$ and 
covering 1\,000\,\AA\ or more in the optical range ($4\,000$--$7\,000$\,\AA). 
These data are easily obtained for the \corot\ targets -- both in the
asteroseismic field and the planet field -- although for the faintest 
planet-hosting stars large telescopes are needed to get the required S/N.
For slowly rotating stars ($v \sin i < 25$\,\kms) the metallicity 
can be determined to $0.05$ dex, provided the atmospheric parameters 
are constrained to better than $100$\,K and $0.1$\,dex for \teff\ and \logg, respectively. 
We will discuss the uncertainties in more detail
and describe our method and results in the next Section.

%
   \begin{figure}
   \centering
\includegraphics[width=9.2cm]{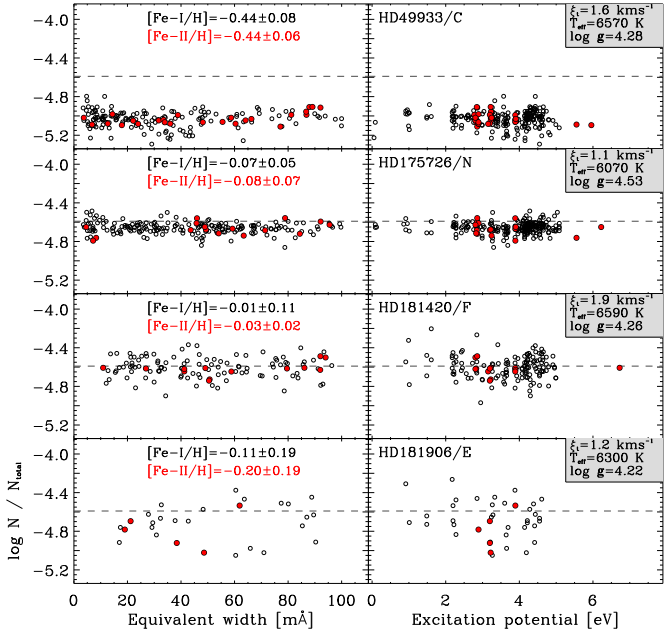}
\vskip 0.2cm
   \caption{Abundance determined for Fe lines versus equivalent width
and excitation potential. Results are shown for the best spectrum
available for the four targets. Open and solid circles are for neutral 
and ionised lines, respectively. 
The dashed lines mark the abundance in the Sun.
              \label{fig:vwares}}
    \end{figure}
%
%

\section{Detailed spectral analysis with \vwa}
\label{sec:vwa}

To analyse the spectra listed in Table~\ref{tab:spec}
we used the software package \vwa, 
which has been described in detail by \cite{bruntt04, bruntt08}.
The software uses atomic data from the \vald\ database
\citep{kupka99} and interpolation of atmospheric models 
either from a modified \atlasni\ grid \citep{heiter02} or
the \marcs\ grid \citep{gustafsson08}. In the current study
we have used the \marcs\ grid which implements the recently
updated abundances in the Sun \citep{grevesse07}.

We first analysed the solar spectrum from \cite{hinkle00}, 
originally published by \cite{kurucz84}.
We used a model atmosphere with $T_{\rm eff} = 5777$\,K, $\log g = 4.44$,
${\rm [Fe/H]} = 0.0$ and $\xi_t=1.0$\,\kms, interpolated in the same grid as used for the four \corot\ stars. 
We determined abundances for 1413 lines in the solar spectrum in the range $3\,777$ to $9\,290$\,\AA\ 
using the VALD oscillator strengths (\loggf). 
We next adjusted the \loggf\ values so each line gives the same abundance as in \cite{grevesse07}. 
Only lines with adjusted \loggf\ values were used in the analysis of the \corot\ targets. 
In Tables~\ref{tab:linelist1}--\ref{tab:linelist3} we list the line parameters (on-line material).
This purely differential analysis was also used by \cite{bruntt08} who discussed the approach in more detail. 
As a test of the robustness of method we analysed three other solar spectra 
obtained with HARPS\footnote{http://www.eso.org/sci/facilities/lasilla/instruments/harps/inst/ monitoring/sun.html} \citep{mayor03} 
which have a resolution, wavelength coverage and S/N comparable to the best \corot\ spectra.
We find that the determined \teff, \logg\ and 
individual abundances are consistent with the Sun within the uncertainties 
(for more details on this analysis see Bruntt et al.\ 2009, in preparation).

In the first step of the analysis, 
\vwa\ automatically identifies isolated lines 
in the spectrum and fits them iteratively using the initial
values of \teff, \logg, and metallicity from the photometric indices.
By analysing the correlations of neutral Fe lines with
equivalent width (EW) and excitation potential (EP) the
\teff\ and \vmicro\ are automatically adjusted.
Similarly, \vwa\ requires that abundances of neutral and ionised
Fe lines give the same 
abundance\footnote{In the following we call this the ``ionisation balance'',
which is the difference in abundance: 
$A$(Fe\,\ito) $-A$(Fe\,\iet)},
which is adjusted by changing primarily \logg. 
Changes in \teff\ also modify the ionisation balance, 
\ie\ the resulting \teff\ and \logg\ are correlated. 
For the stars we have analysed an adjustment of \teff\ of $+100$\,K 
can be compensated by a change in \logg\ of $+0.1$ dex. However, this
degeneration is partially lifted when the quality of the data is good.

To evaluate the uncertainties of the atmospheric parameters we 
repeat the analysis while perturbing \teff, \logg\ and \vmicro.
From this we can determine when the correlations with EW and EP become
significant or the ionisation balance begins to deviate
(for details see \citealt{bruntt08}). 
We emphasise that this way of evaluating the uncertainties
gives only an internal estimate since the absolute temperature 
scale of the model atmospheres may be systematically wrong. 
In other words, our results may indicate a good precision of \teff\ and \logg, 
but the accuracy is most likely not as good. To evaluate the true accuracy
we are currently undertaking the analysis of stars (Bruntt et al.\ 2009, in preparation)
for which \teff\ and \logg\ are determined by methods that are only weakly model dependent
(using interferometry and modelling of eclipsing binary stars).
The preliminary results from this investigation indicate that 
the uncertainties given in current work are realistic.

%
\begin{table}
\caption{Atmospheric parameters determined with \vwa.
The first column lists the HD number and the spectrum used
(\cf\ Table~\ref{tab:spec}). 
In the analysis of HD~181906 \logg\ could not be determined.
\label{tab:vwa}}

\centering          
\setlength{\tabcolsep}{3.5pt}
\begin{tabular}{r@{}l | r@{}r l@{}r l@{}l l@{}l} 
\hline\hline       


HD & /Sp. & \multicolumn{2}{c}{\teff\ [K]} & \multicolumn{2}{c}{\logg} & 
            \multicolumn{2}{c}{\feh} &  \multicolumn{2}{c}{\vmicro\ [km/s]}  \\

\hline 

 49933&/H-A  & $6560\pm$&$110  $&$  4.28$&$\pm0.05 $&$-0.44$  &$\pm0.05  $ & $ 1.67$&$ \pm0.06$ \\ 
 49933&/H-C  & $6570\pm$&$ 70  $&$  4.28$&$\pm0.05 $&$-0.44$  &$\pm0.04  $ & $ 1.60$&$ \pm0.10$ \\ \hline
      
175726&/N & $6070\pm$ &$ 45  $& $ 4.53$&$\pm0.04$ &$-0.07$  &$\pm0.03  $ & $ 1.10$&$\pm0.05$ \\
175726&/E & $6000\pm$ &$ 85  $& $ 4.29$&$\pm0.10 $&$-0.09$  &$\pm0.05  $ & $ 1.07$&$\pm0.11 $ \\ \hline 
      
181420&/F  & $6590\pm$&$ 130  $&$ 4.26$&$\pm0.07 $&$-0.01$  &$\pm0.08  $ & $ 1.91$&$\pm0.15$ \\
181420&/E  & $6565\pm$&$ 180  $&$ 4.27$&$\pm0.10 $&$+0.02$  &$\pm0.11  $ & $ 1.69$&$\pm0.10$ \\ \hline
      
181906&/E & $6300\pm$&$150  $&$     $&$$        &$-0.11$  &$\pm0.14  $ & $1.20$&$\pm0.15$ \\

\hline                    
 \hline                  


\end{tabular}
\end{table}
%

\cite{rent96} have investigated the effects of 
non-LTE on Fe abundances which is important for stars earlier than F type,
especially for stars more metal poor than the Sun. 
Her calculations show that
Fe\,\iet\ is affected while Fe\,\ito\ is nearly unaffected.
As a consequence, the ionisation balance is shifted
relative to results from our applied LTE atmosphere models and
it is therefore necessary to make adjustments in \logg\ \citep{bruntt08}.
We refer to \cite{collet05} (and references therein) 
for a recent discussion of the effects of UV line blocking on non-LTE calculations for Fe.
Extrapolating from Figs.~4 and 5 in \cite{rent96} we find 
a correction of $+0.05$\,dex for Fe\,\iet\ in HD~49933, 
which is the hottest and most metal poor star in our sample.
HD~181420 has a \teff\ similar to HD~49933 but it 
is more metal rich, so we estimate the correction to be $+0.03$\,dex.
The adjustment of \logg\ needed to get ionisation 
balance is about twice these corrections 
($\Delta \log g = +0.10$ and $+0.06$\,dex, respectively).
Since we have used extrapolations of the work
by \cite{rent96} this leads to an additional uncertainty on \logg. 
We have therefore added $0.05$ to the uncertainty on \logg\ from \vwa\
for HD~49933 and HD~181420.

In Fig.~\ref{fig:vwares} we show the abundances of Fe in 
the best spectrum for each of the four stars.
The abundances are shown versus equivalent width and excitation potential.
Open circles are Fe\,\iet\ and solid circles are Fe\,\ito\ lines.
The horizontal dashed lines mark the solar abundance. The mean abundance
and RMS scatter of Fe\,\iet\ and Fe\,\ito\ lines are given in each panel.
The results for the other spectra listed in Table~\ref{tab:spec} give
similar results, but the overall scatter is larger. 

We have checked whether the Fe abundances are correlated with other
atomic line parameters. For HD~175726 and HD~181906 we find a
significant correlation with wavelength. 
For HD~175726 this is especially true for the \elodie\ spectrum, 
while it is less evident in the \narval\ spectrum.
This could be an indication of problems with 
the subtraction of scattered light during the 
reduction of the \elodie\ spectrum. For HD~181906 there 
is a strong correlation of Fe\,\iet\ and wavelength.
From close inspection of the observed and synthetic spectra 
we see a clear asymmetry in the observed line profiles
due to the contamination of light from another star. 
We shall discuss the detailed analysis of this star in Sect.~\ref{sec:bin}.

The atmospheric parameters we determine 
for each spectrum are given in Table~\ref{tab:vwa}. 
For our final result we have computed the weighted mean values and
they are listed in Table~\ref{tab:comp}. 
For the reason stated above the result for the HD~175726/E spectrum was rejected.
The \logg\ determined from spectroscopy is systematically
$0.1$ dex higher than determined from the \shotgun\ method,
and we will investigate this in future work.
In Table~\ref{tab:comp} we also compare our results with 
values found in the literature. There is quite good agreement except for a few cases.
For HD~49933 \cite{kallinger09} found $\log g = 3.9\pm0.1$, 
which is significantly lower than the other determinations. 
For HD~175726 \cite{gillon06} found a significantly higher \teff\ and \logg. 
They used the same \elodie\ spectrum that we rejected. 
Also, as we have discussed, \teff\ and \logg\ are correlated 
and this could be an example of this problem 
(see also Sect.~6.2 in \citealt{bruntt08}).

The abundances in the four \corot\ targets are 
given in Table~\ref{tab:abund} and the abundance pattern
is shown in Fig.~\ref{fig:abund}.
For each of the 17 elements the result is given from left to right for
HD~49933, HD~175726, HD~181420 and HD~181906. The open and solid circles
are the mean abundances determined from neutral and ionised lines, respectively.
To be able to show the results on the same scale we plot all abundances
relative to the Fe\,\iet\ abundance in each star. 
The uncertainty on the abundances includes contributions from the error
on the mean value and the contribution from the uncertainty on the atmospheric parameters. 
When only one or two lines are available we assume an uncertainty of $0.1$\,dex.
We have not included hyperfine structure levels for Mn 
and this probably explains the apparent underabundance of this element.

%
\begin{table}
\caption{Fundamental parameters from this study and the literature.
\label{tab:comp}}

\centering          
\setlength{\tabcolsep}{2.5pt}
\begin{tabular}{r|l@{}l|l@{}l|l@{}l|l}
\hline\hline       
\multicolumn{1}{c|}{HD}   & \multicolumn{2}{c|}{\teff} & \multicolumn{2}{c|}{\logg} & \multicolumn{2}{c|}{\feh} & Source  \\ 
\hline 

 49933 &  $6570$&$\pm 60$ & $4.28 $&$\pm0.06 $ & $-0.44 $&$\pm0.03 $ & \vwa     \\
       &  $6630$&$\pm 90$ & $4.30 $&$\pm0.15 $ & $-0.48 $&$\pm0.12 $ & \str     \\
       &        &         & $4.189$&$\pm0.035$ &         &           & \shotgun \\
       &  $6450$&$\pm 75$ & $3.9  $&$\pm0.1  $ & $      $&$        $ & Kallinger (2009) \\
       &  $6780$&$\pm130$ & $4.24 $&$\pm0.13 $ & $-0.46 $&$\pm0.08 $ & Bruntt (2008) \\
       &  $6598$&$\pm 60$ & $4.08 $&$\pm0.2  $ & $-0.29 $&$\pm0.1  $ & \cite{cenarro07} \\ 
       &  $6735$&$\pm 53$ & $4.26 $&$\pm0.08 $ & $-0.37 $&$\pm0.03 $ & Gillon (2006) \\ 
       &  $6780$&$\pm 70$ & $4.3  $&$\pm0.2  $ & $-0.30 $&$\pm0.11 $ & Bruntt (2004) \\ \hline

175726 &  $6070$&$\pm 45$ & $4.53$ &$\pm0.04 $ & $-0.07 $&$\pm0.03 $ & \vwa     \\
       &  $5980$&$\pm 90$ & $4.68 $&$\pm0.15 $ & $-0.21 $&$\pm0.10 $ & \str     \\
       &        &         & $4.424$&$\pm0.040$ &         &           & \shotgun \\
       &  $5998$&$\pm 44$ & $4.41 $&$\pm0.06 $ & $-0.10 $&$\pm0.03 $ & Valenti (2005) \\
       &  $6036$&$\pm 15$ & $4.40 $&$        $ & $-0.12 $&$        $ & Kovtyukh (2004) \\
       &  $6217$&$\pm 32$ & $4.61 $&$\pm0.04 $ & $+0.03 $&$\pm0.03 $ & Gillon (2006) \\ \hline

181420 &  $6580$&$\pm105$ & $4.26 $&$\pm0.08 $ & $+0.00 $&$\pm0.06 $ & \vwa     \\
       &  $6660$&$\pm 90$ & $4.16 $&$\pm0.15 $ & $-0.07 $&$\pm0.11 $ & \str     \\
       &        &         & $4.151$&$\pm0.027$ &         &           & \shotgun \\ \hline

181906 &  $6300$&$\pm150$ &        &           & $-0.11 $&$\pm0.14 $ & \vwa     \\
       &  $6430$&$\pm 90$ & $4.31 $&$\pm0.15 $ & $-0.20 $&$\pm0.11 $ & \str     \\
       &        &         & $4.220$&$\pm0.056$ &         &           & \shotgun \\

\hline \hline                  
\end{tabular}
\end{table}
%


Abundance analyses of two of the stars are found in the literature.
\cite{valenti05} published a homogeneous set of 
fundamental parameters for 1,040 FGK-type stars
with spectra from radial velocity surveys to find exoplanets. 
They used synthetic spectra to determine \teff, \logg, and
abundances of Na, Si, Ti, Fe and Ni. Only HD~175726 is
included in their study and they find the star to have
an overall metallicity of ${\rm [M/H]}=-0.10$. 
Our results agree remarkably well for
the individual six elements with the largest difference being $0.03$\,dex.
\cite{gillon06} analysed HD~49933 and HD~175726. Our results are
in acceptable agreement for HD~49933 while for HD~175726 the
results disagree. The differences can be explained by 
the higher \teff\ and \logg\ determined by \cite{gillon06}.
Finally, \cite{kallinger09} also analysed \harps\ spectra of
HD~49933 but they found evidence that only Fe is underabundant
while carbon and oxygen are close to the solar value.
Our abundance for carbon is based only two lines, 
but does not confirm this result.





\begin{table}
 \centering
 \caption{Abundances relative to the Sun for the four \corot\ targets.
 \label{tab:abund}}
 \setlength{\tabcolsep}{1pt} 
 \begin{footnotesize}
\begin{tabular}{l|r@{}l r |r@{}l r| r@{}l r|r@{}l r}
\hline \hline
           & \multicolumn{3}{c|}{HD~49933/H-C} & \multicolumn{3}{c|}{ HD~175726/N} & \multicolumn{3}{c|}{ HD~181420/F} & \multicolumn{3}{c}{ HD~181906}  \\
\hline

  {C  \sc   i} &  $ -0.58$&$        $  &   2 &  $ -0.26$&$\pm0.14 $ &   3 &  $ -0.13$&$\pm0.25 $ &   4 &  $      $&$         $&      \\ 
  {O  \sc   i} &  $      $&$        $  &     &  $ -0.09$&$\pm0.03 $ &   3 &  $ +0.15$&$        $ &   2 &  $      $&$         $&      \\ 
  {Na \sc   i} &  $ -0.40$&$\pm0.06 $  &   3 &  $ -0.19$&$\pm0.03 $ &   3 &  $ +0.17$&$        $ &   2 &  $      $&$         $&      \\ 
  {Mg \sc   i} &  $ -0.47$&$        $  &   1 &  $ -0.06$&$        $ &   1 &  $ -0.16$&$        $ &   2 &  $      $&$         $&      \\ 
  {Al \sc   i} &  $      $&$        $  &     &  $ -0.13$&$        $ &   1 &  $ +0.09$&$        $ &   2 &  $      $&$         $&      \\ 
  {Si \sc   i} &  $ -0.43$&$\pm0.06 $  &  10 &  $ -0.13$&$\pm0.03 $ &  16 &  $ -0.04$&$\pm0.08 $ &   7 &  $      $&$         $&      \\ 
  {Si \sc  ii} &  $ -0.35$&$\pm0.05 $  &   2 &  $      $&$        $ &     &  $ +0.10$&$        $ &   1 &  $      $&$         $&      \\ 
  {S  \sc   i} &  $ -0.40$&$        $  &   1 &  $ -0.14$&$        $ &   1 &  $ +0.00$&$        $ &   2 &  $      $&$         $&      \\ 
  {Ca \sc   i} &  $ -0.39$&$\pm0.06 $  &  11 &  $ -0.01$&$\pm0.03 $ &   3 &  $ +0.02$&$\pm0.08 $ &  12 &  $ +0.14$&$\pm0.10  $&   4  \\ 
  {Sc \sc  ii} &  $ -0.46$&$\pm0.06 $  &   5 &  $ -0.12$&$\pm0.04 $ &   5 &  $ +0.01$&$\pm0.09 $ &   4 &  $      $&$         $&      \\ 
  {Ti \sc   i} &  $ -0.37$&$\pm0.06 $  &  13 &  $ -0.04$&$\pm0.04 $ &  13 &  $ +0.08$&$        $ &   2 &  $      $&$         $&      \\ 
  {Ti \sc  ii} &  $ -0.40$&$\pm0.06 $  &  11 &  $ -0.09$&$\pm0.03 $ &  12 &  $ -0.15$&$\pm0.10 $ &   3 &  $ -0.21$&$         $&   1  \\ 
  {V  \sc   i} &  $ -0.53$&$        $  &   1 &  $ -0.07$&$\pm0.04 $ &   8 &  $      $&$        $ &     &  $      $&$         $&      \\ 
  {Cr \sc   i} &  $ -0.49$&$\pm0.06 $  &  12 &  $ -0.09$&$\pm0.03 $ &  18 &  $ -0.04$&$\pm0.08 $ &   5 &  $ -0.19$&$         $&   2  \\ 
  {Cr \sc  ii} &  $ -0.44$&$\pm0.09 $  &   3 &  $ -0.08$&$\pm0.06 $ &   4 &  $ -0.19$&$        $ &   1 &  $ -0.18$&$\pm0.12  $&   4  \\ 
  {Mn \sc   i} &  $ -0.88$&$\pm0.07 $  &   9 &  $ -0.26$&$\pm0.03 $ &   4 &  $ -0.35$&$\pm0.11 $ &   3 &  $ -0.28$&$\pm0.24  $&   3  \\ 
  {Fe \sc   i} &  $ -0.49$&$\pm0.05 $  & 193 &  $ -0.08$&$\pm0.03 $ & 232 &  $ -0.05$&$\pm0.07 $ & 141 &  $ -0.10$&$\pm0.10  $&  37  \\ 
  {Fe \sc  ii} &  $ -0.43$&$\pm0.06 $  &  23 &  $ -0.05$&$\pm0.03 $ &  15 &  $ +0.00$&$\pm0.07 $ &  18 &  $ -0.18$&$\pm0.12  $&   6  \\ 
  {Co \sc   i} &  $ -0.50$&$        $  &   1 &  $ -0.20$&$        $ &   1 &  $      $&$        $ &     &  $      $&$         $&      \\ 
  {Ni \sc   i} &  $ -0.52$&$\pm0.06 $  &  29 &  $ -0.19$&$\pm0.03 $ &  46 &  $ -0.10$&$\pm0.08 $ &  28 &  $ -0.43$&$\pm0.12  $&   8  \\ 
  {Y  \sc  ii} &  $ -0.47$&$\pm0.07 $  &   2 &  $ +0.03$&$        $ &   1 &  $ +0.01$&$\pm0.28 $ &   2 &  $      $&$         $&      \\ 
 
\hline \hline
\end{tabular}
\end{footnotesize}
\end{table}


   \begin{figure}
   \centering
   \includegraphics[angle=90,width=9.1cm]{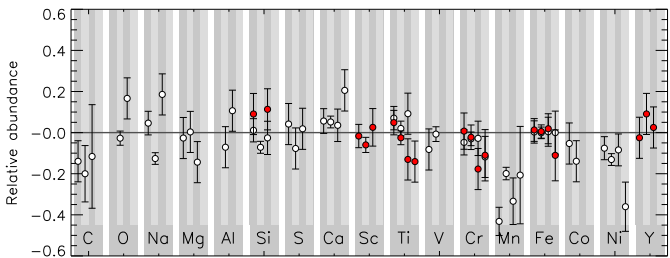}
   \caption{Abundance pattern in the four target stars for 17 elements.
For each element the result is given from left to right for 
HD~49933, 175726, 181420 and 181906. 
Open circles are the mean abundances determined 
from neutral lines and solid circles are for singly ionised lines.
To show the patterns on the same
scale all elements are offset relative to the 
abundance of Fe\,\iet.
              \label{fig:abund}}
    \end{figure}
%
%

%
   \begin{figure*}
   \centering
   \includegraphics[width=16.1cm]{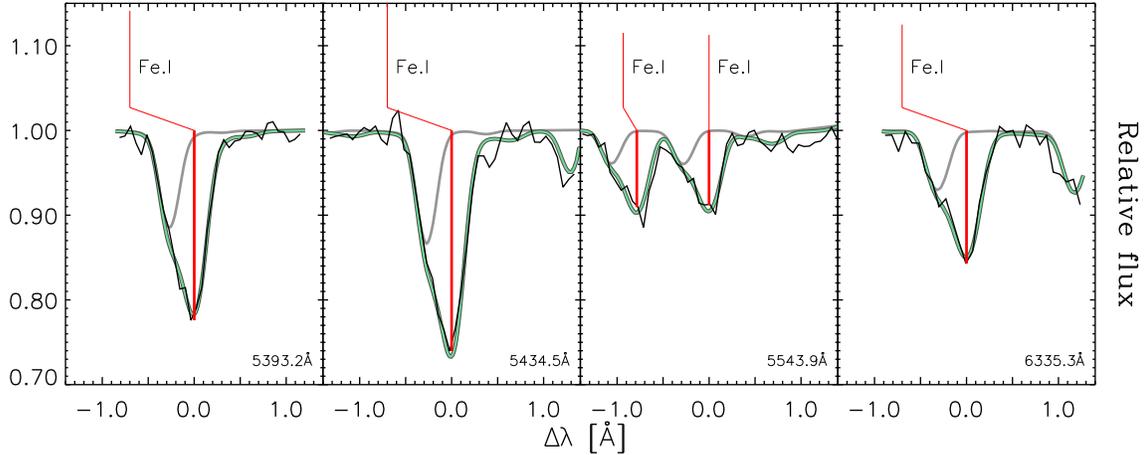}
   \caption{Examples of four Fe\,\iet\ lines fitted by \vwa\ for HD~181906. 
The observed lines (black) are asymmetric due to
the contribution from a fainter star in the blue wing of the lines.
The contribution to the spectrum from the 
contaminating star is taken into account in the analysis.
The synthetic spectrum for the faint star is shown in grey and the
combined spectrum is shown in green.
              \label{fig:fit181906}}
    \end{figure*}
%
%

\section{The binary nature of HD~181906\label{sec:bin}}

Examples of lines fitted by \vwa\ for HD~181906 are shown in Fig.~\ref{fig:fit181906}.
The observed profiles show an asymmetry due to the contamination of a fainter star
in the blue wing of the absorption lines.
We have recently expanded \vwa\ to be able to analyse binary stars,
where each component has a different atmospheric model, line list and chemical composition \citep{clausen08}. 
We initially assumed the composite spectrum was a binary since \cite{makarov05}
and \cite{frankowski07} have found strong evidence for this based on \hipp\ astrometric data.

Due to the relatively relatively weak lines in the fainter star,
we assumed a low luminosity ratio ($R_L=L_{\rm p}/L_{\rm s}=0.3$ at $6\,000$\,\AA).
For the secondary we initially assumed $T_{\rm eff}=5\,500$\,K and $\log g=4.4$
and for the primary we used \teff\ from the $V-K$ index and \logg\ from \shotgun.
The values of \logg\ were fixed throughout the analysis since not enough 
usable Fe\,\ito\ lines were available. 
\vwa\ takes into account the change in $R_L$ with wavelength 
while assuming Planck curves are valid representations
of the variation of the flux with wavelength. We computed abundances for
a grid for different assumptions of $R_L$ and \teff\ of the two components.
Unfortunately, the S/N of the spectrum is quite poor and few lines are available.
Therefore, the parameters of the secondary star cannot be determined very accurately. 
The best solution was found by minimising the
correlations of Fe\,\iet\ with EW and EP (as for single stars) and the
ionisation balance (only for the primary),
and with the additional requirement that both stars have the same Fe abundance.
We found $R_L=0.50\pm0.15$ and for the secondary $T_{\rm eff}=6500\pm250$\,K. 
Since the secondary is fainter but apparently a hotter star, 
we speculate that it is likely a background star rather than 
a part of a binary system. If this is true
our assumption that the two stars have the same metallicity breaks down.

HD~181906 has the lowest power per oscillation mode of the four
\corot\ stars analysed here \citep{michel08}. 
Our discovery that another relatively bright star is within
the photometric aperture may partially explain this low amplitude. 
To investigate this possibility it 
is important to further constrain the parameters of the two stars,
so we will acquire new spectra with high S/N.

%

\section{Conclusion}

We have determined the fundamental parameters of 
four solar-like stars observed with \corot.
This work provides important input for the 
detailed theoretical modelling of the stars based
on comparison with individual observed oscillation frequencies.

We described the \shotgun\ software from which
the mass, radius and age are estimated by comparison
with theoretical evolution tracks. The uncertainty
on the mass is $5$--$10$\% and for the radius $2$--$4$\%.
We can thus constrain \logg\ better than $0.06$\,dex.

The luminosity was estimated using the updated parallaxes
from \cite{leeuwen07} and the uncertainty lies in the range $5$--$13$\%.
We determined \teff\ from
photometric indices and find excellent agreement
with results from our detailed abundance analysis using high-S/N spectra. 
For the two stars with the highest S/N spectra (HD~49933 and HD~175726)
we can constrain \teff\ to within 60\,K, which is better than through photometric indices.
This is not the case for the two stars with relatively poor spectra (HD~181420 and HD~181906).
We have determined the metallicity of the stars and the
abundance pattern for up to 17 elements. 
Within the uncertainties the abundance pattern
can be scaled from the Sun through the metallicity.

We found that the spectrum of HD~181906 consists of two components.
From our analysis of the composite spectrum we obtain a 
luminosity ratio of $R_L=0.50\pm0.15$.
However, the fainter star appears to be hotter than the bright component.
We therefore think the faint star may be a background star rather 
than being physically bound in a binary system. 
Further observations are needed to confirm this result.


%

\begin{acknowledgements}
This project was supported by the Australian and Danish Research Councils.
We made use of the SIMBAD database, operated at CDS, Strasbourg, France.
We used data from GAUDI, the data archive and access system of the ground-based 
asteroseismology programme of the \corot\ mission. The GAUDI system is maintained 
at LAEFF. LAEFF is part of the Space Science Division of INTA.
We used atomic data extracted from the VALD data base 
made available through the Institute of Astronomy in Vienna, Austria. 
We thank C.\ Catala for providing the \narval\ spectrum of HD~175726.

%



\end{acknowledgements}

\bibliographystyle{aa}
\bibliography{bruntt_corotsymp} 

\appendix

\section{Lists of lines used in the abundance analysis}

   \begin{table*}
      \caption[]{The atomic number, element name, 
wavelength, and \loggf\ from the \vald\ database and the adjusted \loggf\ value.
The letters a--d indicate in which spectra the line was used in the analysis:
a $=$ HD~49933/H-C, b $=$ HD~175726/N, c $=$ 181420/F and d $=$ HD181906/E.
         \label{tab:linelist1}}
\centering                          
\begin{tiny} 
\begin{tabular}{r@{\hskip 0.25cm}c@{\hskip 0.25cm}c@{\hskip 0.25cm}c@{\hskip 0.25cm}l|r@{\hskip 0.25cm}c@{\hskip 0.25cm}c@{\hskip 0.25cm}c@{\hskip 0.25cm}l|r@{\hskip 0.25cm}c@{\hskip 0.25cm}c@{\hskip 0.25cm}c@{\hskip 0.25cm}l}
\hline\hline

 & & \vald & Adjusted & &
 & & \vald & Adjusted & &
 & & \vald & Adjusted & \\

 \eeee & $\lambda$ [\AA] & \loggf & \loggf & Spectra &
 \eeee & $\lambda$ [\AA] & \loggf & \loggf & Spectra &
 \eeee & $\lambda$ [\AA] & \loggf & \loggf & Spectra \\

\hline

 $^{  6}$C\ione & 4932.049 & $ -1.884$ & $ -1.038$ & abc  &  $$     Sc\itwo & 6245.637 & $ -1.030$ & $ -1.053$ & ab   &  $$     Cr\itwo & 5305.853 & $ -2.357$ & $ -1.902$ & bd   \\
  $$     C\ione & 5380.337 & $ -1.842$ & $ -1.452$ & a    &  $$     Sc\itwo & 6604.601 & $ -1.309$ & $ -1.258$ & abc  &  $$     Cr\itwo & 5308.408 & $ -1.846$ & $ -1.704$ & d    \\
  $$     C\ione & 5800.602 & $ -2.338$ & $ -2.165$ & c    & $^{ 22}$Ti\ione & 4465.805 & $ -0.163$ & $ -0.250$ & a    &  $$     Cr\itwo & 5313.563 & $ -1.650$ & $ -1.465$ & bd   \\
  $$     C\ione & 7113.179 & $ -0.774$ & $ -0.658$ & bc   &  $$     Ti\ione & 4512.734 & $ -0.480$ & $ -0.475$ & ab   & $^{ 25}$Mn\ione & 4709.712 & $ -0.340$ & $ -0.252$ & a    \\
  $$     C\ione & 7116.988 & $ -0.907$ & $ -0.658$ & bc   &  $$     Ti\ione & 4534.776 & $+ 0.280$ & $+ 0.225$ & b    &  $$     Mn\ione & 4754.042 & $ -0.086$ & $+ 0.358$ & abcd \\
 $^{  8}$O\ione & 7771.941 & $+ 0.369$ & $+ 0.784$ & b    &  $$     Ti\ione & 4548.763 & $ -0.354$ & $ -0.401$ & a    &  $$     Mn\ione & 4761.512 & $ -0.138$ & $ -0.079$ & a    \\
  $$     O\ione & 7774.161 & $+ 0.223$ & $+ 0.627$ & bc   &  $$     Ti\ione & 4617.269 & $+ 0.389$ & $+ 0.312$ & b    &  $$     Mn\ione & 4762.367 & $+ 0.425$ & $+ 0.571$ & b    \\
  $$     O\ione & 7775.390 & $+ 0.001$ & $+ 0.355$ & bc   &  $$     Ti\ione & 4623.097 & $+ 0.110$ & $+ 0.079$ & b    &  $$     Mn\ione & 4766.418 & $+ 0.100$ & $+ 0.306$ & ab   \\
$^{ 11}$Na\ione & 5682.633 & $ -0.700$ & $ -0.578$ & a    &  $$     Ti\ione & 4758.118 & $+ 0.425$ & $+ 0.358$ & ab   &  $$     Mn\ione & 4783.427 & $+ 0.042$ & $+ 0.434$ & abcd \\
 $$     Na\ione & 5688.205 & $ -0.450$ & $ -0.366$ & abc  &  $$     Ti\ione & 4759.270 & $+ 0.514$ & $+ 0.445$ & c    &  $$     Mn\ione & 5377.637 & $ -0.109$ & $+ 0.179$ & a    \\
 $$     Na\ione & 6154.226 & $ -1.560$ & $ -1.448$ & bc   &  $$     Ti\ione & 4820.411 & $ -0.441$ & $ -0.498$ & ab   &  $$     Mn\ione & 5420.355 & $ -1.462$ & $ -0.934$ & ab   \\
 $$     Na\ione & 6160.747 & $ -1.260$ & $ -1.105$ & ab   &  $$     Ti\ione & 4928.336 & $+ 0.050$ & $ -0.160$ & a    &  $$     Mn\ione & 5537.760 & $ -2.017$ & $ -1.904$ & a    \\
$^{ 12}$Mg\ione & 4571.096 & $ -5.691$ & $ -5.885$ & b    &  $$     Ti\ione & 4981.731 & $+ 0.504$ & $+ 0.429$ & a    &  $$     Mn\ione & 6013.513 & $ -0.251$ & $+ 0.057$ & b    \\
 $$     Mg\ione & 5711.088 & $ -1.833$ & $ -1.647$ & ac   &  $$     Ti\ione & 5016.161 & $ -0.574$ & $ -0.530$ & b    &  $$     Mn\ione & 6021.819 & $+ 0.034$ & $+ 0.213$ & acd  \\
 $$     Mg\ione & 8717.825 & $ -0.930$ & $ -0.886$ & c    &  $$     Ti\ione & 5192.969 & $ -1.006$ & $ -1.038$ & a    & $^{ 26}$Fe\ione & 4080.209 & $ -1.220$ & $ -1.285$ & a    \\
$^{ 13}$Al\ione & 6696.023 & $ -1.347$ & $ -1.549$ & c    &  $$     Ti\ione & 5210.385 & $ -0.884$ & $ -0.900$ & ab   &  $$     Fe\ione & 4080.877 & $ -1.800$ & $ -1.827$ & a    \\
 $$     Al\ione & 6698.673 & $ -1.647$ & $ -1.807$ & b    &  $$     Ti\ione & 5426.250 & $ -3.006$ & $ -3.024$ & b    &  $$     Fe\ione & 4120.207 & $ -1.267$ & $ -1.222$ & a    \\
 $$     Al\ione & 8772.865 & $ -0.316$ & $ -0.233$ & c    &  $$     Ti\ione & 5866.451 & $ -0.840$ & $ -0.929$ & a    &  $$     Fe\ione & 4136.521 & $ -1.516$ & $ -1.456$ & a    \\
$^{ 14}$Si\ione & 5517.533 & $ -2.384$ & $ -2.429$ & ab   &  $$     Ti\ione & 6258.102 & $ -0.355$ & $ -0.411$ & ab   &  $$     Fe\ione & 4136.998 & $ -0.453$ & $ -0.608$ & a    \\
 $$     Si\ione & 5645.613 & $ -2.140$ & $ -2.005$ & a    &  $$     Ti\ione & 6258.706 & $ -0.240$ & $ -0.329$ & ac   &  $$     Fe\ione & 4139.927 & $ -3.629$ & $ -3.506$ & a    \\
 $$     Si\ione & 5666.677 & $ -1.050$ & $ -1.598$ & a    &  $$     Ti\ione & 6261.098 & $ -0.479$ & $ -0.512$ & ab   &  $$     Fe\ione & 4168.942 & $ -1.650$ & $ -1.695$ & b    \\
 $$     Si\ione & 5675.417 & $ -1.030$ & $ -1.039$ & c    &  $$     Ti\ione & 8434.957 & $ -0.886$ & $ -0.802$ & b    &  $$     Fe\ione & 4184.892 & $ -0.869$ & $ -0.851$ & ab   \\
 $$     Si\ione & 5684.484 & $ -1.650$ & $ -1.551$ & b    &  $$     Ti\ione & 8435.652 & $ -1.023$ & $ -1.031$ & b    &  $$     Fe\ione & 4199.095 & $+ 0.155$ & $+ 0.029$ & a    \\
 $$     Si\ione & 5690.425 & $ -1.870$ & $ -1.827$ & b    &  $$     Ti\itwo & 4316.799 & $ -1.580$ & $ -1.437$ & b    &  $$     Fe\ione & 4365.897 & $ -2.250$ & $ -2.267$ & ab   \\
 $$     Si\ione & 5708.400 & $ -1.470$ & $ -1.348$ & a    &  $$     Ti\itwo & 4411.925 & $ -2.550$ & $ -2.305$ & ab   &  $$     Fe\ione & 4389.245 & $ -4.583$ & $ -4.522$ & a    \\
 $$     Si\ione & 5747.667 & $ -0.780$ & $ -1.404$ & b    &  $$     Ti\itwo & 4417.719 & $ -1.230$ & $ -0.900$ & abc  &  $$     Fe\ione & 4432.568 & $ -1.600$ & $ -1.736$ & b    \\
 $$     Si\ione & 5753.623 & $ -0.830$ & $ -1.265$ & a    &  $$     Ti\itwo & 4470.857 & $ -2.060$ & $ -1.895$ & b    &  $$     Fe\ione & 4433.782 & $ -1.267$ & $ -1.249$ & c    \\
 $$     Si\ione & 5793.073 & $ -2.060$ & $ -1.936$ & b    &  $$     Ti\itwo & 4493.513 & $ -2.830$ & $ -2.775$ & a    &  $$     Fe\ione & 4438.343 & $ -1.630$ & $ -1.679$ & b    \\
 $$     Si\ione & 5948.541 & $ -1.230$ & $ -1.083$ & b    &  $$     Ti\itwo & 4544.028 & $ -2.530$ & $ -2.518$ & ab   &  $$     Fe\ione & 4439.881 & $ -3.002$ & $ -3.007$ & ab   \\
 $$     Si\ione & 6091.919 & $ -1.400$ & $ -1.251$ & c    &  $$     Ti\itwo & 4589.958 & $ -1.620$ & $ -1.446$ & abc  &  $$     Fe\ione & 4445.471 & $ -5.441$ & $ -5.455$ & ab   \\
 $$     Si\ione & 6138.515 & $ -1.350$ & $ -1.283$ & a    &  $$     Ti\itwo & 4779.985 & $ -1.260$ & $ -1.147$ & ab   &  $$     Fe\ione & 4480.137 & $ -1.933$ & $ -1.818$ & b    \\
 $$     Si\ione & 6142.483 & $ -0.920$ & $ -1.423$ & ab   &  $$     Ti\itwo & 5129.152 & $ -1.300$ & $ -1.005$ & b    &  $$     Fe\ione & 4485.676 & $ -1.020$ & $ -1.096$ & abc  \\
 $$     Si\ione & 6145.016 & $ -0.820$ & $ -1.356$ & bc   &  $$     Ti\itwo & 5211.536 & $ -1.356$ & $ -1.369$ & ac   &  $$     Fe\ione & 4489.739 & $ -3.966$ & $ -3.951$ & b    \\
 $$     Si\ione & 6155.134 & $ -0.400$ & $ -0.684$ & a    &  $$     Ti\itwo & 5336.771 & $ -1.630$ & $ -1.476$ & b    &  $$     Fe\ione & 4492.678 & $ -1.650$ & $ -1.627$ & b    \\
 $$     Si\ione & 6237.319 & $ -0.530$ & $ -0.967$ & ac   &  $$     Ti\itwo & 5381.015 & $ -1.970$ & $ -1.897$ & a    &  $$     Fe\ione & 4542.412 & $ -2.050$ & $ -1.753$ & c    \\
 $$     Si\ione & 6244.466 & $ -0.690$ & $ -1.222$ & b    &  $$     Ti\itwo & 5418.751 & $ -2.110$ & $ -1.997$ & abd  &  $$     Fe\ione & 4547.847 & $ -1.012$ & $ -0.927$ & abc  \\
 $$     Si\ione & 6414.980 & $ -1.100$ & $ -0.992$ & bc   &  $$     Ti\itwo & 5490.690 & $ -2.650$ & $ -2.660$ & a    &  $$     Fe\ione & 4551.647 & $ -2.060$ & $ -1.928$ & b    \\
 $$     Si\ione & 6527.202 & $ -1.500$ & $ -1.110$ & b    &  $$     Ti\itwo & 6491.561 & $ -1.793$ & $ -2.032$ & ab   &  $$     Fe\ione & 4587.128 & $ -1.737$ & $ -1.671$ & abc  \\
 $$     Si\ione & 6635.687 & $ -1.630$ & $ -1.816$ & a    &  $$     Ti\itwo & 7575.423 & $ -1.397$ & $ -1.327$ & b    &  $$     Fe\ione & 4595.359 & $ -1.758$ & $ -1.612$ & b    \\
 $$     Si\ione & 6800.596 & $ -1.640$ & $ -1.650$ & b    &  $^{ 23}$V\ione & 5627.633 & $ -0.363$ & $ -0.588$ & b    &  $$     Fe\ione & 4602.001 & $ -3.154$ & $ -3.197$ & abc  \\
 $$     Si\ione & 7405.772 & $ -0.820$ & $ -0.632$ & b    &   $$     V\ione & 5670.853 & $ -0.420$ & $ -0.571$ & b    &  $$     Fe\ione & 4602.941 & $ -2.209$ & $ -2.188$ & abcd \\
 $$     Si\ione & 7423.496 & $ -0.314$ & $ -0.520$ & bc   &   $$     V\ione & 5727.048 & $ -0.012$ & $ -0.137$ & b    &  $$     Fe\ione & 4607.647 & $ -1.545$ & $ -1.210$ & ac   \\
 $$     Si\ione & 7918.384 & $ -0.610$ & $ -0.409$ & b    &   $$     V\ione & 6039.722 & $ -0.650$ & $ -0.810$ & b    &  $$     Fe\ione & 4625.045 & $ -1.340$ & $ -1.396$ & abc  \\
 $$     Si\ione & 7932.348 & $ -0.470$ & $ -0.216$ & b    &   $$     V\ione & 6090.214 & $ -0.062$ & $ -0.268$ & b    &  $$     Fe\ione & 4630.120 & $ -2.587$ & $ -2.561$ & ab   \\
 $$     Si\ione & 7944.001 & $ -0.310$ & $ -0.056$ & c    &   $$     V\ione & 6119.523 & $ -0.320$ & $ -0.566$ & b    &  $$     Fe\ione & 4635.846 & $ -2.358$ & $ -2.349$ & ab   \\
 $$     Si\ione & 8443.970 & $ -1.400$ & $ -1.285$ & b    &   $$     V\ione & 6199.197 & $ -1.300$ & $ -1.606$ & b    &  $$     Fe\ione & 4637.503 & $ -1.390$ & $ -1.285$ & ab   \\
 $$     Si\itwo & 6347.109 & $+ 0.297$ & $+ 0.506$ & a    &   $$     V\ione & 6243.105 & $ -0.980$ & $ -1.117$ & ab   &  $$     Fe\ione & 4638.010 & $ -1.119$ & $ -1.057$ & ab   \\
 $$     Si\itwo & 6371.371 & $ -0.003$ & $+ 0.152$ & ac   & $^{ 24}$Cr\ione & 4319.636 & $ -0.820$ & $ -1.094$ & b    &  $$     Fe\ione & 4661.534 & $ -1.270$ & $ -1.119$ & ab   \\
 $^{ 16}$S\ione & 6052.674 & $ -0.740$ & $ -0.483$ & c    &  $$     Cr\ione & 4545.945 & $ -1.370$ & $ -1.446$ & b    &  $$     Fe\ione & 4661.970 & $ -2.502$ & $ -2.361$ & b    \\
  $$     S\ione & 6757.171 & $ -0.310$ & $ -0.208$ & abc  &  $$     Cr\ione & 4616.120 & $ -1.190$ & $ -1.334$ & ab   &  $$     Fe\ione & 4669.171 & $ -1.211$ & $ -1.156$ & b    \\
$^{ 20}$Ca\ione & 4425.437 & $ -0.286$ & $ -0.491$ & ac   &  $$     Cr\ione & 4626.174 & $ -1.320$ & $ -1.414$ & ab   &  $$     Fe\ione & 4690.138 & $ -1.645$ & $ -1.500$ & b    \\
 $$     Ca\ione & 4578.551 & $ -0.170$ & $ -0.562$ & abd  &  $$     Cr\ione & 4651.282 & $ -1.460$ & $ -1.496$ & abc  &  $$     Fe\ione & 4728.546 & $ -1.172$ & $ -0.927$ & acd  \\
 $$     Ca\ione & 5349.465 & $ -1.178$ & $ -0.256$ & b    &  $$     Cr\ione & 4652.152 & $ -1.030$ & $ -0.916$ & abd  &  $$     Fe\ione & 4733.592 & $ -2.988$ & $ -2.998$ & abcd \\
 $$     Ca\ione & 5581.965 & $ -0.569$ & $ -0.607$ & abcd &  $$     Cr\ione & 4801.047 & $ -0.131$ & $ -0.179$ & b    &  $$     Fe\ione & 4736.773 & $ -0.752$ & $ -0.701$ & a    \\
 $$     Ca\ione & 5588.749 & $+ 0.313$ & $+ 0.296$ & c    &  $$     Cr\ione & 4936.335 & $ -0.340$ & $ -0.245$ & ab   &  $$     Fe\ione & 4741.530 & $ -1.765$ & $ -1.940$ & ab   \\
 $$     Ca\ione & 5867.562 & $ -0.801$ & $ -1.634$ & a    &  $$     Cr\ione & 5247.566 & $ -1.640$ & $ -1.623$ & ab   &  $$     Fe\ione & 4745.800 & $ -1.270$ & $ -1.102$ & abc  \\
 $$     Ca\ione & 6122.217 & $ -0.386$ & $ -0.347$ & ac   &  $$     Cr\ione & 5287.200 & $ -0.907$ & $ -0.921$ & b    &  $$     Fe\ione & 4757.582 & $ -2.321$ & $ -1.958$ & b    \\
 $$     Ca\ione & 6166.439 & $ -1.156$ & $ -1.161$ & bcd  &  $$     Cr\ione & 5296.691 & $ -1.400$ & $ -1.375$ & abd  &  $$     Fe\ione & 4779.439 & $ -2.020$ & $ -2.198$ & ab   \\
 $$     Ca\ione & 6169.042 & $ -0.804$ & $ -0.814$ & a    &  $$     Cr\ione & 5297.376 & $+ 0.167$ & $ -0.014$ & abc  &  $$     Fe\ione & 4785.957 & $ -1.930$ & $ -1.748$ & b    \\
 $$     Ca\ione & 6169.563 & $ -0.527$ & $ -0.522$ & ac   &  $$     Cr\ione & 5300.744 & $ -2.120$ & $ -2.153$ & bc   &  $$     Fe\ione & 4787.827 & $ -2.530$ & $ -2.578$ & abc  \\
 $$     Ca\ione & 6439.075 & $+ 0.394$ & $+ 0.370$ & cd   &  $$     Cr\ione & 5329.142 & $ -0.064$ & $ -0.063$ & b    &  $$     Fe\ione & 4788.757 & $ -1.763$ & $ -1.770$ & ab   \\
 $$     Ca\ione & 6449.808 & $ -1.015$ & $ -0.480$ & a    &  $$     Cr\ione & 5348.312 & $ -1.290$ & $ -1.238$ & abc  &  $$     Fe\ione & 4798.265 & $ -1.174$ & $ -1.378$ & b    \\
 $$     Ca\ione & 6455.598 & $ -1.557$ & $ -1.348$ & a    &  $$     Cr\ione & 5783.093 & $ -0.500$ & $ -0.455$ & ab   &  $$     Fe\ione & 4799.406 & $ -2.230$ & $ -2.115$ & a    \\
 $$     Ca\ione & 6471.662 & $ -0.653$ & $ -0.668$ & cd   &  $$     Cr\ione & 5783.886 & $ -0.295$ & $ -0.271$ & ab   &  $$     Fe\ione & 4802.880 & $ -1.514$ & $ -1.519$ & bcd  \\
 $$     Ca\ione & 6493.781 & $+ 0.019$ & $ -0.107$ & bc   &  $$     Cr\ione & 5787.965 & $ -0.083$ & $ -0.258$ & a    &  $$     Fe\ione & 4807.709 & $ -2.200$ & $ -2.016$ & b    \\
 $$     Ca\ione & 6499.650 & $ -0.719$ & $ -0.787$ & ac   &  $$     Cr\ione & 6661.078 & $ -0.190$ & $ -0.165$ & b    &  $$     Fe\ione & 4809.938 & $ -2.720$ & $ -2.550$ & a    \\
 $$     Ca\ione & 6717.681 & $ -0.596$ & $ -0.474$ & acd  &  $$     Cr\ione & 7400.226 & $ -0.111$ & $ -0.161$ & c    &  $$     Fe\ione & 4839.544 & $ -1.822$ & $ -1.713$ & bd   \\
 $$     Ca\ione & 7326.145 & $+ 0.073$ & $ -0.049$ & c    &  $$     Cr\itwo & 4554.988 & $ -1.282$ & $ -1.356$ & a    &  $$     Fe\ione & 4886.326 & $ -0.556$ & $ -0.547$ & ab   \\
$^{ 21}$Sc\itwo & 5239.813 & $ -0.765$ & $ -0.672$ & abc  &  $$     Cr\itwo & 4616.629 & $ -1.361$ & $ -1.160$ & ab   &  $$     Fe\ione & 4892.859 & $ -1.290$ & $ -1.182$ & acd  \\
 $$     Sc\itwo & 5526.790 & $+ 0.024$ & $+ 0.069$ & abc  &  $$     Cr\itwo & 4824.127 & $ -0.970$ & $ -0.759$ & b    &  $$     Fe\ione & 4917.230 & $ -1.180$ & $ -0.945$ & ab   \\
 $$     Sc\itwo & 5669.042 & $ -1.200$ & $ -1.072$ & abc  &  $$     Cr\itwo & 5237.329 & $ -1.160$ & $ -1.103$ & acd  &  $$     Fe\ione & 4924.770 & $ -2.241$ & $ -2.085$ & ac   \\

\hline
\end{tabular}
\end{tiny}
\end{table*}

   \begin{table*}
      \caption[]{The atomic number, element name, 
wavelength, and \loggf\ from the \vald\ database and the adjusted \loggf\ value.
The letters a--d indicate in which spectra the line was used in the analysis:
a $=$ HD~49933/H-C, b $=$ HD~175726/N, c $=$ 181420/F and d $=$ HD181906/E.
         \label{tab:linelist2}}
\centering                          

\begin{tiny} 
\begin{tabular}{r@{\hskip 0.25cm}c@{\hskip 0.25cm}c@{\hskip 0.25cm}c@{\hskip 0.25cm}l|r@{\hskip 0.25cm}c@{\hskip 0.25cm}c@{\hskip 0.25cm}c@{\hskip 0.25cm}l|r@{\hskip 0.25cm}c@{\hskip 0.25cm}c@{\hskip 0.25cm}c@{\hskip 0.25cm}l}
\hline\hline

 & & \vald & Adjusted & &
 & & \vald & Adjusted & &
 & & \vald & Adjusted & \\

 \eeee & $\lambda$ [\AA] & \loggf & \loggf & Spectra &
 \eeee & $\lambda$ [\AA] & \loggf & \loggf & Spectra &
 \eeee & $\lambda$ [\AA] & \loggf & \loggf & Spectra \\
\hline

$^{ 26}$Fe\ione & 4930.315 & $ -1.201$ & $ -0.956$ & abc  &  $$     Fe\ione & 5400.502 & $ -0.160$ & $ -0.212$ & ab   &  $$     Fe\ione & 5856.088 & $ -1.328$ & $ -1.582$ & abc  \\
 $$     Fe\ione & 4946.388 & $ -1.170$ & $ -1.140$ & abc  &  $$     Fe\ione & 5401.269 & $ -1.920$ & $ -1.701$ & ab   &  $$     Fe\ione & 5858.778 & $ -2.260$ & $ -2.294$ & b    \\
 $$     Fe\ione & 4950.106 & $ -1.670$ & $ -1.490$ & a    &  $$     Fe\ione & 5405.775 & $ -1.844$ & $ -1.961$ & a    &  $$     Fe\ione & 5859.578 & $ -0.398$ & $ -0.482$ & ac   \\
 $$     Fe\ione & 4962.572 & $ -1.182$ & $ -1.098$ & abc  &  $$     Fe\ione & 5406.775 & $ -1.720$ & $ -1.410$ & b    &  $$     Fe\ione & 5861.110 & $ -2.450$ & $ -2.402$ & b    \\
 $$     Fe\ione & 4966.089 & $ -0.871$ & $ -0.890$ & abcd &  $$     Fe\ione & 5409.134 & $ -1.300$ & $ -1.037$ & ab   &  $$     Fe\ione & 5862.353 & $ -0.058$ & $ -0.158$ & ab   \\
 $$     Fe\ione & 4967.890 & $ -0.622$ & $ -0.471$ & abd  &  $$     Fe\ione & 5415.199 & $+ 0.642$ & $+ 0.427$ & acd  &  $$     Fe\ione & 5905.672 & $ -0.730$ & $ -0.736$ & c    \\
 $$     Fe\ione & 4969.918 & $ -0.710$ & $ -0.745$ & ac   &  $$     Fe\ione & 5424.068 & $+ 0.520$ & $+ 0.559$ & a    &  $$     Fe\ione & 5916.247 & $ -2.994$ & $ -2.956$ & bc   \\
 $$     Fe\ione & 4970.496 & $ -1.740$ & $ -1.655$ & b    &  $$     Fe\ione & 5434.524 & $ -2.122$ & $ -2.243$ & ad   &  $$     Fe\ione & 5927.789 & $ -1.090$ & $ -1.092$ & b    \\
 $$     Fe\ione & 4973.102 & $ -0.950$ & $ -0.821$ & abc  &  $$     Fe\ione & 5445.042 & $ -0.020$ & $ -0.034$ & acd  &  $$     Fe\ione & 5929.677 & $ -1.410$ & $ -1.149$ & b    \\
 $$     Fe\ione & 4977.647 & $ -2.153$ & $ -1.958$ & ab   &  $$     Fe\ione & 5464.280 & $ -1.402$ & $ -1.564$ & acd  &  $$     Fe\ione & 5930.180 & $ -0.230$ & $ -0.221$ & ac   \\
 $$     Fe\ione & 4986.223 & $ -1.390$ & $ -1.269$ & ab   &  $$     Fe\ione & 5466.396 & $ -0.630$ & $ -0.611$ & abc  &  $$     Fe\ione & 5933.800 & $ -2.230$ & $ -2.132$ & b    \\
 $$     Fe\ione & 4988.950 & $ -0.890$ & $ -0.644$ & ab   &  $$     Fe\ione & 5466.988 & $ -2.233$ & $ -2.273$ & ab   &  $$     Fe\ione & 5934.655 & $ -1.170$ & $ -1.132$ & abc  \\
 $$     Fe\ione & 4994.130 & $ -3.080$ & $ -3.187$ & abcd &  $$     Fe\ione & 5470.094 & $ -1.810$ & $ -1.565$ & b    &  $$     Fe\ione & 5952.718 & $ -1.440$ & $ -1.345$ & b    \\
 $$     Fe\ione & 5014.943 & $ -0.303$ & $ -0.302$ & b    &  $$     Fe\ione & 5472.709 & $ -1.495$ & $ -1.500$ & ac   &  $$     Fe\ione & 5956.694 & $ -4.605$ & $ -4.652$ & b    \\
 $$     Fe\ione & 5029.618 & $ -2.050$ & $ -1.953$ & c    &  $$     Fe\ione & 5473.900 & $ -0.760$ & $ -0.728$ & abc  &  $$     Fe\ione & 5976.775 & $ -1.310$ & $ -1.131$ & c    \\
 $$     Fe\ione & 5044.211 & $ -2.038$ & $ -2.089$ & bd   &  $$     Fe\ione & 5480.861 & $ -1.260$ & $ -1.145$ & b    &  $$     Fe\ione & 5983.673 & $ -1.878$ & $ -0.493$ & abcd \\
 $$     Fe\ione & 5049.820 & $ -1.355$ & $ -1.390$ & cd   &  $$     Fe\ione & 5481.243 & $ -1.243$ & $ -1.257$ & b    &  $$     Fe\ione & 5984.814 & $ -0.343$ & $ -0.045$ & b    \\
 $$     Fe\ione & 5054.643 & $ -1.921$ & $ -1.983$ & a    &  $$     Fe\ione & 5487.145 & $ -1.530$ & $ -1.375$ & ab   &  $$     Fe\ione & 5987.066 & $ -0.556$ & $ -0.283$ & bc   \\
 $$     Fe\ione & 5068.766 & $ -1.042$ & $ -1.186$ & b    &  $$     Fe\ione & 5497.516 & $ -2.849$ & $ -2.996$ & b    &  $$     Fe\ione & 6003.012 & $ -1.120$ & $ -1.030$ & abcd \\
 $$     Fe\ione & 5072.668 & $ -1.224$ & $ -0.815$ & b    &  $$     Fe\ione & 5506.779 & $ -2.797$ & $ -2.918$ & ab   &  $$     Fe\ione & 6005.543 & $ -3.192$ & $ -3.496$ & b    \\
 $$     Fe\ione & 5076.262 & $ -0.767$ & $ -0.872$ & a    &  $$     Fe\ione & 5522.447 & $ -1.550$ & $ -1.412$ & bc   &  $$     Fe\ione & 6007.960 & $ -0.966$ & $ -0.551$ & a    \\
 $$     Fe\ione & 5083.339 & $ -2.958$ & $ -3.086$ & ab   &  $$     Fe\ione & 5525.544 & $ -1.084$ & $ -1.192$ & ab   &  $$     Fe\ione & 6008.554 & $ -1.078$ & $ -0.764$ & abc  \\
 $$     Fe\ione & 5090.774 & $ -0.400$ & $ -0.440$ & a    &  $$     Fe\ione & 5531.984 & $ -1.610$ & $ -1.321$ & ab   &  $$     Fe\ione & 6024.058 & $ -0.120$ & $+ 0.013$ & abc  \\
 $$     Fe\ione & 5109.652 & $ -0.980$ & $ -0.680$ & ac   &  $$     Fe\ione & 5543.936 & $ -1.140$ & $ -1.016$ & abd  &  $$     Fe\ione & 6027.051 & $ -1.089$ & $ -1.045$ & c    \\
 $$     Fe\ione & 5121.639 & $ -0.810$ & $ -0.772$ & b    &  $$     Fe\ione & 5546.506 & $ -1.310$ & $ -1.050$ & abc  &  $$     Fe\ione & 6034.035 & $ -2.420$ & $ -2.314$ & b    \\
 $$     Fe\ione & 5127.359 & $ -3.307$ & $ -3.346$ & ab   &  $$     Fe\ione & 5553.578 & $ -1.410$ & $ -1.301$ & abc  &  $$     Fe\ione & 6035.338 & $ -2.590$ & $ -2.620$ & b    \\
 $$     Fe\ione & 5131.469 & $ -2.515$ & $ -2.379$ & a    &  $$     Fe\ione & 5560.212 & $ -1.190$ & $ -1.030$ & abd  &  $$     Fe\ione & 6056.005 & $ -0.460$ & $ -0.411$ & bc   \\
 $$     Fe\ione & 5133.689 & $+ 0.140$ & $+ 0.210$ & ad   &  $$     Fe\ione & 5562.706 & $ -0.659$ & $ -0.744$ & a    &  $$     Fe\ione & 6065.482 & $ -1.530$ & $ -1.576$ & ac   \\
 $$     Fe\ione & 5137.382 & $ -0.400$ & $ -0.300$ & b    &  $$     Fe\ione & 5565.704 & $ -0.285$ & $ -0.001$ & ad   &  $$     Fe\ione & 6078.491 & $ -0.424$ & $ -0.108$ & a    \\
 $$     Fe\ione & 5141.739 & $ -1.964$ & $ -2.156$ & abc  &  $$     Fe\ione & 5567.391 & $ -2.564$ & $ -2.770$ & ab   &  $$     Fe\ione & 6079.009 & $ -1.120$ & $ -0.974$ & ab   \\
 $$     Fe\ione & 5145.094 & $ -2.876$ & $ -3.191$ & ab   &  $$     Fe\ione & 5569.618 & $ -0.486$ & $ -0.654$ & a    &  $$     Fe\ione & 6082.711 & $ -3.573$ & $ -3.615$ & b    \\
 $$     Fe\ione & 5150.840 & $ -3.003$ & $ -3.463$ & c    &  $$     Fe\ione & 5572.842 & $ -0.275$ & $ -0.346$ & ac   &  $$     Fe\ione & 6085.259 & $ -3.095$ & $ -2.983$ & ab   \\
 $$     Fe\ione & 5151.911 & $ -3.322$ & $ -3.226$ & abcd &  $$     Fe\ione & 5576.089 & $ -1.000$ & $ -0.936$ & abc  &  $$     Fe\ione & 6093.644 & $ -1.500$ & $ -1.334$ & ab   \\
 $$     Fe\ione & 5159.058 & $ -0.820$ & $ -0.790$ & bcd  &  $$     Fe\ione & 5586.756 & $ -0.120$ & $ -0.183$ & ac   &  $$     Fe\ione & 6094.374 & $ -1.940$ & $ -1.564$ & b    \\
 $$     Fe\ione & 5194.942 & $ -2.090$ & $ -2.279$ & ab   &  $$     Fe\ione & 5608.972 & $ -2.400$ & $ -2.255$ & b    &  $$     Fe\ione & 6096.665 & $ -1.930$ & $ -1.804$ & ab   \\
 $$     Fe\ione & 5195.468 & $ -0.002$ & $ -0.081$ & ab   &  $$     Fe\ione & 5618.633 & $ -1.276$ & $ -1.271$ & a    &  $$     Fe\ione & 6098.245 & $ -1.880$ & $ -1.760$ & ab   \\
 $$     Fe\ione & 5196.077 & $ -0.451$ & $ -0.636$ & a    &  $$     Fe\ione & 5619.595 & $ -1.700$ & $ -1.440$ & abc  &  $$     Fe\ione & 6102.173 & $ -0.627$ & $ -0.070$ & ab   \\
 $$     Fe\ione & 5198.711 & $ -2.135$ & $ -2.175$ & abc  &  $$     Fe\ione & 5624.022 & $ -1.480$ & $ -1.073$ & b    &  $$     Fe\ione & 6105.131 & $ -2.050$ & $ -1.930$ & b    \\
 $$     Fe\ione & 5215.181 & $ -0.871$ & $ -1.036$ & a    &  $$     Fe\ione & 5624.542 & $ -0.755$ & $ -0.827$ & ac   &  $$     Fe\ione & 6127.907 & $ -1.399$ & $ -1.367$ & abc  \\
 $$     Fe\ione & 5217.389 & $ -1.070$ & $ -1.150$ & abcd &  $$     Fe\ione & 5633.947 & $ -0.270$ & $ -0.146$ & abc  &  $$     Fe\ione & 6136.615 & $ -1.400$ & $ -1.476$ & c    \\
 $$     Fe\ione & 5228.377 & $ -1.290$ & $ -1.061$ & c    &  $$     Fe\ione & 5635.823 & $ -1.890$ & $ -1.567$ & abc  &  $$     Fe\ione & 6137.692 & $ -1.403$ & $ -1.437$ & a    \\
 $$     Fe\ione & 5235.387 & $ -0.854$ & $ -0.959$ & b    &  $$     Fe\ione & 5638.262 & $ -0.870$ & $ -0.756$ & ab   &  $$     Fe\ione & 6151.618 & $ -3.299$ & $ -3.338$ & ac   \\
 $$     Fe\ione & 5236.204 & $ -1.497$ & $ -1.563$ & abc  &  $$     Fe\ione & 5641.434 & $ -1.180$ & $ -1.066$ & a    &  $$     Fe\ione & 6157.728 & $ -1.260$ & $ -1.117$ & a    \\
 $$     Fe\ione & 5242.491 & $ -0.967$ & $ -0.953$ & ac   &  $$     Fe\ione & 5650.706 & $ -0.960$ & $ -0.674$ & ab   &  $$     Fe\ione & 6165.360 & $ -1.474$ & $ -1.446$ & abc  \\
 $$     Fe\ione & 5243.777 & $ -1.150$ & $ -1.012$ & b    &  $$     Fe\ione & 5661.346 & $ -1.736$ & $ -1.864$ & a    &  $$     Fe\ione & 6170.507 & $ -0.440$ & $ -0.329$ & ab   \\
 $$     Fe\ione & 5247.050 & $ -4.946$ & $ -4.992$ & b    &  $$     Fe\ione & 5677.685 & $ -2.700$ & $ -2.646$ & b    &  $$     Fe\ione & 6173.336 & $ -2.880$ & $ -2.897$ & abc  \\
 $$     Fe\ione & 5250.646 & $ -2.181$ & $ -2.105$ & ac   &  $$     Fe\ione & 5679.023 & $ -0.920$ & $ -0.717$ & bc   &  $$     Fe\ione & 6180.204 & $ -2.586$ & $ -2.710$ & bc   \\
 $$     Fe\ione & 5253.462 & $ -1.573$ & $ -1.631$ & abc  &  $$     Fe\ione & 5691.497 & $ -1.520$ & $ -1.414$ & ab   &  $$     Fe\ione & 6187.990 & $ -1.720$ & $ -1.644$ & ab   \\
 $$     Fe\ione & 5281.790 & $ -0.834$ & $ -0.995$ & ac   &  $$     Fe\ione & 5696.090 & $ -1.720$ & $ -1.890$ & b    &  $$     Fe\ione & 6200.313 & $ -2.437$ & $ -2.404$ & ab   \\
 $$     Fe\ione & 5283.621 & $ -0.432$ & $ -0.551$ & a    &  $$     Fe\ione & 5701.545 & $ -2.216$ & $ -2.202$ & ab   &  $$     Fe\ione & 6213.430 & $ -2.482$ & $ -2.586$ & abcd \\
 $$     Fe\ione & 5285.129 & $ -1.640$ & $ -1.495$ & ab   &  $$     Fe\ione & 5705.465 & $ -1.355$ & $ -1.463$ & b    &  $$     Fe\ione & 6216.352 & $ -1.425$ & $ -1.380$ & a    \\
 $$     Fe\ione & 5288.525 & $ -1.508$ & $ -1.619$ & abd  &  $$     Fe\ione & 5717.833 & $ -1.130$ & $ -0.993$ & ab   &  $$     Fe\ione & 6219.281 & $ -2.433$ & $ -2.484$ & acd  \\
 $$     Fe\ione & 5293.959 & $ -1.870$ & $ -1.701$ & ab   &  $$     Fe\ione & 5724.455 & $ -2.640$ & $ -2.556$ & b    &  $$     Fe\ione & 6226.736 & $ -2.220$ & $ -2.088$ & b    \\
 $$     Fe\ione & 5295.312 & $ -1.690$ & $ -1.524$ & ab   &  $$     Fe\ione & 5731.762 & $ -1.300$ & $ -1.098$ & b    &  $$     Fe\ione & 6229.228 & $ -2.805$ & $ -2.942$ & ac   \\
 $$     Fe\ione & 5302.302 & $ -0.720$ & $ -0.816$ & d    &  $$     Fe\ione & 5738.228 & $ -2.340$ & $ -2.531$ & b    &  $$     Fe\ione & 6230.723 & $ -1.281$ & $ -1.352$ & ac   \\
 $$     Fe\ione & 5307.361 & $ -2.987$ & $ -3.052$ & bc   &  $$     Fe\ione & 5741.848 & $ -1.854$ & $ -1.638$ & bc   &  $$     Fe\ione & 6232.641 & $ -1.223$ & $ -1.112$ & abc  \\
 $$     Fe\ione & 5315.070 & $ -1.550$ & $ -1.436$ & b    &  $$     Fe\ione & 5752.023 & $ -1.267$ & $ -0.801$ & abd  &  $$     Fe\ione & 6246.319 & $ -0.733$ & $ -0.828$ & acd  \\
 $$     Fe\ione & 5321.108 & $ -0.951$ & $ -1.235$ & b    &  $$     Fe\ione & 5760.345 & $ -2.490$ & $ -2.395$ & b    &  $$     Fe\ione & 6252.555 & $ -1.687$ & $ -1.742$ & ac   \\
 $$     Fe\ione & 5322.041 & $ -2.803$ & $ -2.994$ & bcd  &  $$     Fe\ione & 5762.992 & $ -0.450$ & $ -0.364$ & b    &  $$     Fe\ione & 6265.134 & $ -2.550$ & $ -2.561$ & ac   \\
 $$     Fe\ione & 5329.989 & $ -1.189$ & $ -1.066$ & b    &  $$     Fe\ione & 5775.081 & $ -1.298$ & $ -1.045$ & ab   &  $$     Fe\ione & 6270.225 & $ -2.464$ & $ -2.589$ & b    \\
 $$     Fe\ione & 5361.625 & $ -1.430$ & $ -1.277$ & abcd &  $$     Fe\ione & 5780.600 & $ -2.640$ & $ -2.465$ & ab   &  $$     Fe\ione & 6271.279 & $ -2.703$ & $ -2.743$ & a    \\
 $$     Fe\ione & 5364.871 & $+ 0.228$ & $+ 0.113$ & abc  &  $$     Fe\ione & 5793.915 & $ -1.700$ & $ -1.626$ & b    &  $$     Fe\ione & 6330.850 & $ -1.740$ & $ -1.179$ & abc  \\
 $$     Fe\ione & 5365.399 & $ -1.020$ & $ -1.147$ & a    &  $$     Fe\ione & 5798.171 & $ -1.890$ & $ -1.766$ & ab   &  $$     Fe\ione & 6335.331 & $ -2.177$ & $ -2.336$ & acd  \\
 $$     Fe\ione & 5367.467 & $+ 0.443$ & $+ 0.232$ & ad   &  $$     Fe\ione & 5806.725 & $ -1.050$ & $ -0.872$ & ac   &  $$     Fe\ione & 6336.824 & $ -0.856$ & $ -0.903$ & ac   \\
 $$     Fe\ione & 5373.709 & $ -0.860$ & $ -0.776$ & abc  &  $$     Fe\ione & 5809.218 & $ -1.840$ & $ -1.867$ & ac   &  $$     Fe\ione & 6338.877 & $ -1.060$ & $ -0.870$ & b    \\
 $$     Fe\ione & 5379.574 & $ -1.514$ & $ -1.474$ & ab   &  $$     Fe\ione & 5811.914 & $ -2.430$ & $ -2.430$ & b    &  $$     Fe\ione & 6355.029 & $ -2.350$ & $ -2.125$ & ac   \\
 $$     Fe\ione & 5383.369 & $+ 0.645$ & $+ 0.450$ & acd  &  $$     Fe\ione & 5814.807 & $ -1.970$ & $ -1.858$ & ab   &  $$     Fe\ione & 6380.743 & $ -1.376$ & $ -1.252$ & abc  \\
 $$     Fe\ione & 5386.334 & $ -1.770$ & $ -1.706$ & ab   &  $$     Fe\ione & 5845.287 & $ -1.820$ & $ -1.896$ & b    &  $$     Fe\ione & 6393.601 & $ -1.432$ & $ -1.596$ & ac   \\
 $$     Fe\ione & 5389.479 & $ -0.410$ & $ -0.466$ & abc  &  $$     Fe\ione & 5848.123 & $ -0.903$ & $ -1.184$ & abc  &  $$     Fe\ione & 6400.001 & $ -0.290$ & $ -0.386$ & a    \\
 $$     Fe\ione & 5393.168 & $ -0.715$ & $ -0.821$ & acd  &  $$     Fe\ione & 5852.219 & $ -1.330$ & $ -1.244$ & bc   &  $$     Fe\ione & 6408.018 & $ -1.018$ & $ -0.866$ & a    \\
 $$     Fe\ione & 5398.279 & $ -0.670$ & $ -0.619$ & bc   &  $$     Fe\ione & 5855.077 & $ -1.478$ & $ -1.564$ & b    &  $$     Fe\ione & 6411.649 & $ -0.595$ & $ -0.670$ & ac   \\

\hline
\end{tabular}
\end{tiny}
\end{table*}

    \begin{table*}
      \caption[]{The atomic number, element name, 
wavelength, and \loggf\ from the \vald\ database and the adjusted \loggf\ value.
The letters a--d indicate in which spectra the line was used in the analysis:
a $=$ HD~49933/H-C, b $=$ HD~175726/N, c $=$ 181420/F and d $=$ HD181906/E.
         \label{tab:linelist3}}
\centering                          
\begin{tiny} 
\begin{tabular}{r@{\hskip 0.25cm}c@{\hskip 0.25cm}c@{\hskip 0.25cm}c@{\hskip 0.25cm}l|r@{\hskip 0.25cm}c@{\hskip 0.25cm}c@{\hskip 0.25cm}c@{\hskip 0.25cm}l|r@{\hskip 0.25cm}c@{\hskip 0.25cm}c@{\hskip 0.25cm}c@{\hskip 0.25cm}l}
\hline\hline

 & & \vald & Adjusted & &
 & & \vald & Adjusted & &
 & & \vald & Adjusted & \\

 \eeee & $\lambda$ [\AA] & \loggf & \loggf & Spectra &
 \eeee & $\lambda$ [\AA] & \loggf & \loggf & Spectra &
 \eeee & $\lambda$ [\AA] & \loggf & \loggf & Spectra \\
\hline

$^{ 26}$Fe\ione & 6419.950 & $ -0.240$ & $ -0.266$ & c    &  $$     Fe\ione & 7807.952 & $ -0.697$ & $ -0.461$ & bc   &  $$     Ni\ione & 4829.016 & $ -0.330$ & $ -0.330$ & b    \\
 $$     Fe\ione & 6421.351 & $ -2.027$ & $ -2.173$ & ab   &  $$     Fe\ione & 7832.194 & $+ 0.018$ & $+ 0.202$ & b    &  $$     Ni\ione & 4831.169 & $ -0.320$ & $ -0.388$ & c    \\
 $$     Fe\ione & 6430.846 & $ -2.006$ & $ -2.109$ & abcd &  $$     Fe\ione & 7941.089 & $ -2.286$ & $ -2.476$ & b    &  $$     Ni\ione & 4904.407 & $ -0.170$ & $ -0.234$ & abcd \\
 $$     Fe\ione & 6436.407 & $ -2.460$ & $ -2.350$ & b    &  $$     Fe\ione & 8028.309 & $ -0.794$ & $ -0.677$ & c    &  $$     Ni\ione & 4935.831 & $ -0.350$ & $ -0.317$ & ab   \\
 $$     Fe\ione & 6481.870 & $ -2.984$ & $ -2.958$ & bc   &  $$     Fe\ione & 8046.047 & $ -0.082$ & $+ 0.177$ & b    &  $$     Ni\ione & 4953.200 & $ -0.580$ & $ -0.835$ & ab   \\
 $$     Fe\ione & 6498.939 & $ -4.699$ & $ -4.685$ & ab   &  $$     Fe\ione & 8207.745 & $ -0.987$ & $ -0.843$ & b    &  $$     Ni\ione & 4998.218 & $ -0.690$ & $ -0.845$ & ab   \\
 $$     Fe\ione & 6518.367 & $ -2.460$ & $ -2.560$ & c    &  $$     Fe\ione & 8339.398 & $ -1.421$ & $ -0.366$ & b    &  $$     Ni\ione & 5010.934 & $ -0.870$ & $ -0.866$ & bc   \\
 $$     Fe\ione & 6592.914 & $ -1.473$ & $ -1.601$ & ac   &  $$     Fe\ione & 8365.634 & $ -2.047$ & $ -1.990$ & bc   &  $$     Ni\ione & 5017.568 & $ -0.020$ & $ -0.168$ & a    \\
 $$     Fe\ione & 6593.870 & $ -2.422$ & $ -2.374$ & ac   &  $$     Fe\ione & 8401.404 & $ -3.442$ & $ -3.553$ & b    &  $$     Ni\ione & 5035.357 & $+ 0.290$ & $+ 0.060$ & a    \\
 $$     Fe\ione & 6597.561 & $ -1.070$ & $ -0.906$ & ac   &  $$     Fe\ione & 8439.563 & $ -0.698$ & $ -0.672$ & b    &  $$     Ni\ione & 5081.107 & $+ 0.300$ & $+ 0.184$ & ab   \\
 $$     Fe\ione & 6608.026 & $ -4.030$ & $ -4.005$ & b    &  $$     Fe\ione & 8471.739 & $ -0.863$ & $ -0.941$ & b    &  $$     Ni\ione & 5082.339 & $ -0.540$ & $ -0.535$ & abcd \\
 $$     Fe\ione & 6609.110 & $ -2.692$ & $ -2.648$ & bc   &  $$     Fe\ione & 8514.072 & $ -2.229$ & $ -2.239$ & b    &  $$     Ni\ione & 5084.089 & $+ 0.030$ & $ -0.090$ & b    \\
 $$     Fe\ione & 6627.545 & $ -1.680$ & $ -1.485$ & b    &  $$     Fe\ione & 8582.257 & $ -2.134$ & $ -2.086$ & c    &  $$     Ni\ione & 5094.406 & $ -1.080$ & $ -1.108$ & a    \\
 $$     Fe\ione & 6633.750 & $ -0.799$ & $ -0.721$ & ac   &  $$     Fe\ione & 8611.804 & $ -1.926$ & $ -1.895$ & c    &  $$     Ni\ione & 5099.927 & $ -0.100$ & $ -0.173$ & b    \\
 $$     Fe\ione & 6646.932 & $ -3.990$ & $ -3.988$ & b    &  $$     Fe\ione & 8616.276 & $ -0.405$ & $ -0.846$ & c    &  $$     Ni\ione & 5115.389 & $ -0.110$ & $ -0.148$ & acd  \\
 $$     Fe\ione & 6677.987 & $ -1.418$ & $ -1.455$ & ac   &  $$     Fe\ione & 8621.601 & $ -2.321$ & $ -2.267$ & c    &  $$     Ni\ione & 5146.480 & $ -0.060$ & $+ 0.131$ & b    \\
 $$     Fe\ione & 6703.567 & $ -3.160$ & $ -3.055$ & a    &  $$     Fe\ione & 8688.626 & $ -1.212$ & $ -1.210$ & c    &  $$     Ni\ione & 5155.125 & $ -0.650$ & $ -0.642$ & ab   \\
 $$     Fe\ione & 6705.101 & $ -1.496$ & $ -1.021$ & c    &  $$     Fe\ione & 8699.454 & $ -0.380$ & $ -0.329$ & c    &  $$     Ni\ione & 5155.762 & $+ 0.011$ & $ -0.084$ & abcd \\
 $$     Fe\ione & 6713.745 & $ -1.600$ & $ -1.428$ & b    &  $$     Fe\ione & 8710.392 & $ -0.646$ & $ -0.311$ & c    &  $$     Ni\ione & 5435.855 & $ -2.590$ & $ -2.495$ & b    \\
 $$     Fe\ione & 6715.383 & $ -1.640$ & $ -1.436$ & ac   &  $$     Fe\ione & 8838.429 & $ -2.050$ & $ -1.938$ & c    &  $$     Ni\ione & 5593.733 & $ -0.840$ & $ -0.773$ & b    \\
 $$     Fe\ione & 6725.357 & $ -2.300$ & $ -2.186$ & b    &  $$     Fe\itwo & 4416.830 & $ -2.410$ & $ -2.524$ & a    &  $$     Ni\ione & 5614.768 & $ -0.508$ & $ -0.465$ & a    \\
 $$     Fe\ione & 6726.661 & $ -0.829$ & $ -0.976$ & a    &  $$     Fe\itwo & 4491.405 & $ -2.700$ & $ -2.504$ & ab   &  $$     Ni\ione & 5663.975 & $ -0.430$ & $ -0.424$ & ac   \\
 $$     Fe\ione & 6732.065 & $ -2.210$ & $ -2.167$ & b    &  $$     Fe\itwo & 4508.288 & $ -2.250$ & $ -2.333$ & ab   &  $$     Ni\ione & 5682.198 & $ -0.470$ & $ -0.457$ & b    \\
 $$     Fe\ione & 6733.151 & $ -1.580$ & $ -1.424$ & ab   &  $$     Fe\itwo & 4515.339 & $ -2.450$ & $ -2.540$ & ac   &  $$     Ni\ione & 5694.977 & $ -0.610$ & $ -0.626$ & abc  \\
 $$     Fe\ione & 6745.101 & $ -2.160$ & $ -2.106$ & b    &  $$     Fe\itwo & 4520.224 & $ -2.600$ & $ -2.475$ & b    &  $$     Ni\ione & 5754.655 & $ -2.330$ & $ -2.028$ & ad   \\
 $$     Fe\ione & 6750.153 & $ -2.621$ & $ -2.629$ & bc   &  $$     Fe\itwo & 4541.524 & $ -2.790$ & $ -3.002$ & bc   &  $$     Ni\ione & 5805.213 & $ -0.640$ & $ -0.627$ & abc  \\
 $$     Fe\ione & 6752.707 & $ -1.204$ & $ -1.199$ & bc   &  $$     Fe\itwo & 4576.340 & $ -2.920$ & $ -2.901$ & ab   &  $$     Ni\ione & 5996.727 & $ -1.060$ & $ -1.026$ & b    \\
 $$     Fe\ione & 6804.001 & $ -1.496$ & $ -1.503$ & b    &  $$     Fe\itwo & 4620.521 & $ -3.240$ & $ -3.199$ & abc  &  $$     Ni\ione & 6007.306 & $ -3.330$ & $ -3.406$ & b    \\
 $$     Fe\ione & 6806.845 & $ -3.210$ & $ -3.124$ & a    &  $$     Fe\itwo & 4993.358 & $ -3.640$ & $ -3.600$ & ac   &  $$     Ni\ione & 6053.679 & $ -1.070$ & $ -1.019$ & b    \\
 $$     Fe\ione & 6810.263 & $ -0.986$ & $ -0.962$ & ac   &  $$     Fe\itwo & 5132.669 & $ -3.980$ & $ -3.917$ & c    &  $$     Ni\ione & 6086.276 & $ -0.530$ & $ -0.467$ & ac   \\
 $$     Fe\ione & 6820.372 & $ -1.320$ & $ -1.125$ & abc  &  $$     Fe\itwo & 5197.577 & $ -2.100$ & $ -2.185$ & ac   &  $$     Ni\ione & 6108.107 & $ -2.450$ & $ -2.567$ & ab   \\
 $$     Fe\ione & 6828.591 & $ -0.920$ & $ -0.821$ & b    &  $$     Fe\itwo & 5234.625 & $ -2.230$ & $ -2.178$ & ac   &  $$     Ni\ione & 6111.066 & $ -0.870$ & $ -0.845$ & abc  \\
 $$     Fe\ione & 6841.339 & $ -0.750$ & $ -0.617$ & b    &  $$     Fe\itwo & 5256.938 & $ -4.250$ & $ -4.055$ & a    &  $$     Ni\ione & 6175.360 & $ -0.559$ & $ -0.519$ & bc   \\
 $$     Fe\ione & 6843.656 & $ -0.930$ & $ -0.821$ & ab   &  $$     Fe\itwo & 5284.109 & $ -2.990$ & $ -2.955$ & a    &  $$     Ni\ione & 6176.807 & $ -0.260$ & $ -0.265$ & abcd \\
 $$     Fe\ione & 6999.884 & $ -1.560$ & $ -1.386$ & b    &  $$     Fe\itwo & 5325.553 & $ -3.120$ & $ -3.050$ & bcd  &  $$     Ni\ione & 6204.600 & $ -1.100$ & $ -1.125$ & abc  \\
 $$     Fe\ione & 7038.223 & $ -1.300$ & $ -1.164$ & b    &  $$     Fe\itwo & 5362.869 & $ -2.739$ & $ -2.341$ & a    &  $$     Ni\ione & 6223.981 & $ -0.910$ & $ -0.936$ & ac   \\
 $$     Fe\ione & 7068.410 & $ -1.380$ & $ -1.324$ & bc   &  $$     Fe\itwo & 5414.073 & $ -3.540$ & $ -3.447$ & ac   &  $$     Ni\ione & 6259.592 & $ -1.400$ & $ -1.220$ & b    \\
 $$     Fe\ione & 7086.722 & $ -2.682$ & $ -2.497$ & b    &  $$     Fe\itwo & 5425.257 & $ -3.160$ & $ -3.126$ & abcd &  $$     Ni\ione & 6378.247 & $ -0.830$ & $ -0.825$ & ab   \\
 $$     Fe\ione & 7090.383 & $ -1.210$ & $ -1.079$ & bc   &  $$     Fe\itwo & 5427.826 & $ -1.664$ & $ -1.347$ & c    &  $$     Ni\ione & 6482.796 & $ -2.630$ & $ -2.825$ & a    \\
 $$     Fe\ione & 7130.922 & $ -0.790$ & $ -0.721$ & c    &  $$     Fe\itwo & 5534.847 & $ -2.730$ & $ -2.748$ & ab   &  $$     Ni\ione & 6635.118 & $ -0.820$ & $ -0.734$ & ab   \\
 $$     Fe\ione & 7132.986 & $ -1.628$ & $ -1.607$ & bc   &  $$     Fe\itwo & 5991.376 & $ -3.540$ & $ -3.444$ & ac   &  $$     Ni\ione & 6643.629 & $ -2.300$ & $ -2.037$ & abcd \\
 $$     Fe\ione & 7142.503 & $ -0.931$ & $ -0.863$ & bc   &  $$     Fe\itwo & 6084.111 & $ -3.780$ & $ -3.724$ & cd   &  $$     Ni\ione & 6767.768 & $ -2.170$ & $ -2.130$ & bc   \\
 $$     Fe\ione & 7145.316 & $ -1.532$ & $ -1.196$ & c    &  $$     Fe\itwo & 6147.741 & $ -2.721$ & $ -2.654$ & c    &  $$     Ni\ione & 6772.313 & $ -0.980$ & $ -0.991$ & abc  \\
 $$     Fe\ione & 7401.685 & $ -1.599$ & $ -1.524$ & c    &  $$     Fe\itwo & 6149.258 & $ -2.720$ & $ -2.633$ & ac   &  $$     Ni\ione & 7030.006 & $ -1.860$ & $ -1.765$ & b    \\
 $$     Fe\ione & 7411.154 & $ -0.428$ & $ -0.214$ & bc   &  $$     Fe\itwo & 6238.392 & $ -2.630$ & $ -2.765$ & ab   &  $$     Ni\ione & 7110.892 & $ -2.980$ & $ -2.935$ & b    \\
 $$     Fe\ione & 7418.667 & $ -1.376$ & $ -1.382$ & bc   &  $$     Fe\itwo & 6239.366 & $ -4.538$ & $ -4.745$ & a    &  $$     Ni\ione & 7122.191 & $+ 0.040$ & $ -0.245$ & bc   \\
 $$     Fe\ione & 7445.746 & $ -0.237$ & $+ 0.022$ & b    &  $$     Fe\itwo & 6247.557 & $ -2.310$ & $ -2.194$ & abcd &  $$     Ni\ione & 7385.236 & $ -1.970$ & $ -1.935$ & b    \\
 $$     Fe\ione & 7453.998 & $ -2.410$ & $ -2.294$ & b    &  $$     Fe\itwo & 6369.462 & $ -4.160$ & $ -4.052$ & ad   &  $$     Ni\ione & 7393.600 & $ -0.825$ & $ -0.126$ & bc   \\
 $$     Fe\ione & 7461.521 & $ -3.580$ & $ -3.472$ & b    &  $$     Fe\itwo & 6432.680 & $ -3.520$ & $ -3.485$ & ab   &  $$     Ni\ione & 7422.277 & $ -0.140$ & $ -0.277$ & c    \\
 $$     Fe\ione & 7481.933 & $ -1.800$ & $ -1.689$ & b    &  $$     Fe\itwo & 6446.410 & $ -1.960$ & $ -1.885$ & b    &  $$     Ni\ione & 7525.111 & $ -0.546$ & $ -0.582$ & bc   \\
 $$     Fe\ione & 7484.297 & $ -1.700$ & $ -1.553$ & b    &  $$     Fe\itwo & 6456.383 & $ -2.100$ & $ -1.990$ & a    &  $$     Ni\ione & 7555.598 & $ -0.046$ & $+ 0.061$ & bc   \\
 $$     Fe\ione & 7491.649 & $ -1.014$ & $ -0.897$ & bc   &  $$     Fe\itwo & 7711.723 & $ -2.500$ & $ -2.427$ & b    &  $$     Ni\ione & 7714.314 & $ -2.200$ & $ -1.718$ & b    \\
 $$     Fe\ione & 7568.894 & $ -0.882$ & $ -0.774$ & bc   & $^{ 27}$Co\ione & 5352.045 & $+ 0.060$ & $ -0.029$ & a    &  $$     Ni\ione & 7727.613 & $ -0.170$ & $ -0.344$ & bc   \\
 $$     Fe\ione & 7583.788 & $ -1.885$ & $ -1.925$ & b    &  $$     Co\ione & 5647.234 & $ -1.560$ & $ -1.670$ & b    &  $$     Ni\ione & 7748.891 & $ -0.343$ & $ -0.163$ & b    \\
 $$     Fe\ione & 7586.014 & $ -0.871$ & $ -0.026$ & bc   & $^{ 28}$Ni\ione & 4331.640 & $ -2.100$ & $ -2.076$ & b    &  $$     Ni\ione & 7788.936 & $ -2.420$ & $ -1.953$ & bc   \\
 $$     Fe\ione & 7710.364 & $ -1.113$ & $ -1.071$ & b    &  $$     Ni\ione & 4410.512 & $ -1.080$ & $ -1.015$ & b    &  $$     Ni\ione & 7797.586 & $ -0.262$ & $ -0.178$ & bc   \\
 $$     Fe\ione & 7748.269 & $ -1.751$ & $ -1.708$ & b    &  $$     Ni\ione & 4715.757 & $ -0.320$ & $ -0.456$ & abcd &  $^{ 39}$Y\itwo & 4883.684 & $+ 0.070$ & $+ 0.163$ & a    \\
 $$     Fe\ione & 7751.137 & $ -0.895$ & $ -0.667$ & b    &  $$     Ni\ione & 4756.510 & $ -0.270$ & $ -0.325$ & b    &   $$     Y\itwo & 5087.416 & $ -0.170$ & $ -0.196$ & abc  \\
 $$     Fe\ione & 7780.552 & $ -2.361$ & $+ 0.374$ & b    &  $$     Ni\ione & 4806.984 & $ -0.640$ & $ -0.585$ & abc  &   $$     Y\itwo & 5200.406 & $ -0.570$ & $ -0.792$ & c    \\

\hline
\end{tabular}
\end{tiny}
\end{table*}

\end{document}